\documentclass[twocolumn,aps,prl,superscriptaddress]{revtex4-2}
\usepackage{amsmath}
\usepackage{graphicx}
\usepackage{MnSymbol}
\usepackage{hyperref}
\usepackage{fancyhdr}
\usepackage{parskip}

\begin{document}

\title{On how to determine surface roughness power spectra}

\author{N. Rodriguez}
\affiliation{BD Medical-Pharmaceutical Systems 1 Becton Drive, Franklin Lakes, NJ 07419, USA}
\affiliation{3843 Nottingham Loop, Wildwood, FL 32163, USA}

\author{L. Gontard}
\affiliation{BD Medical-Pharmaceutical Systems, 11 Rue Aritides-Verges, Le Pont de Claix 38801, France}

\author{C. Ma}
\affiliation{State Key Laboratory of Solid Lubrication, Lanzhou Institute of Chemical Physics, Chinese Academy of Sciences, 730000 Lanzhou, China}

\author{R. Xu}
\affiliation{State Key Laboratory of Solid Lubrication, Lanzhou Institute of Chemical Physics, Chinese Academy of Sciences, 730000 Lanzhou, China}
\affiliation{Peter Gr\"unberg Institute (PGI-1), Forschungszentrum J\"ulich, 52425, J\"ulich, Germany}
\affiliation{MultiscaleConsulting, Wolfshovener str. 2, 52428 J\"ulich, Germany}

\author{B.N.J. Persson}
\affiliation{Peter Gr\"unberg Institute (PGI-1), Forschungszentrum J\"ulich, 52425, J\"ulich, Germany}
\affiliation{State Key Laboratory of Solid Lubrication, Lanzhou Institute of Chemical Physics, Chinese Academy of Sciences, 730000 Lanzhou, China}
\affiliation{MultiscaleConsulting, Wolfshovener str. 2, 52428 J\"ulich, Germany}

\begin{abstract}
{\bf Abstract}: 
Analytical contact mechanics theories depend on surface roughness through the surface roughness power spectrum. In the present study, we evaluated the usability of various experimental methods for studying surface roughness. Our findings indicated that height data obtained from optical methods often lack accuracy and should not be utilized for calculating surface roughness power spectra. Conversely, engineering stylus instruments and atomic force microscopy (AFM) typically yield reliable results that are consistent across the overlapping roughness length scale region. For surfaces with isotropic roughness, the two-dimensional (2D) power spectrum can be derived from the one-dimensional (1D) power spectrum using several approaches, which we explored in this paper.
\end{abstract}

\maketitle

\thispagestyle{fancy}


{\bf 1 Introduction}

The quality of the surfaces of solids has become extremely important in the design and
production of components in particular in high 
tech applications \cite{MRS1,MRS2,MRS3,MRS4,MRS5} and biological and medical
applications \cite{Nic,Gorb,Gorb1,Gorb2,Gorb3}. 
This is particularly true for the microgeometry (surface roughness). 
The most complete information about
surface roughness is the height probability
distribution $P_h$ and the surface roughness power spectrum $C({\bf q})$,
and most standard surface roughness parameters 
can be obtained from these functions \cite{my,add1,Rev,add2,add3,fluctuate}. 
For randomly rough surfaces the power spectra contain all the
(ensemble averaged) information about the surface
roughness. For this reason
the only information about the surface roughness which enters analytical contact mechanics
theories (with or without adhesion) is the function $C({\bf q})$.
Thus, the  (ensemble averaged) area of real contact, the interfacial stress distribution, and the
distribution of interfacial separations are all determined by
$C({\bf q})$ \cite{Persson2,Prodanov,Carbone1}.

All topography measurements involve interaction between the sensing probe and the surface.
Some methods involve a direct (probe-tip--substrate) solid contact while others involve exposing the studied solid
to electromagnetic waves or electron beams, which may modify the surface topography. 
For solid contact methods like stylus, the contact stress at the probe tip may be very
high resulting in elastoplastic deformation and scratches on the studied surfaces. 
Non-contact methods like optical (e.g., laser-based) methods may locally increase the temperature
and induce thermally activated processes and surface modifications. We note that some materials, like 
soft gels and some biological materials, cannot be studied directly using engineering stylus instruments, 
and even non-contact AFM may be problematic in some cases because it exposes the substrate surfaces to attractive 
(e.g., Van der Waals or electric) forces and repulsive forces when the tip is close to the surface.
For stylus measurements on soft solids, e.g., soft rubber materials, the attractive tip-substrate force can 
effectively modify the surface topography as observed for silicone rubber (PDMS)
where stick-slip occurred \cite{Julia}. 
However, sometimes replicas made using an elastically stiff (glassy) polymer can be obtained from soft (and hard)
solids and the topography of the replica can be studied using stylus instruments \cite{Julia}.
When transparent materials are studied using optical methods several
problems may occur e.g., reflections of the electromagnetic waves from internal surfaces of the sample,
as observed in some cases for ice \cite{Vienna}. Scanning tunneling microscopy (STM) can only be performed
on materials with high enough electric conductivity and the tunneling electrons can result in surface modifications,
at least on the atomic scale.

Several experimental methods have been used to obtain surface roughness power spectra:

(a) Optical methods, e.g., light scattering or interferometric methods \cite{optical}.

(b) Stylus methods, e.g., engineering stylus and atomic force microscopy (AFM) \cite{AFM}.

(c) Optical, Xerographic, X-ray, and neutron reflectivity, or electron microscopy study of the line profile of 
vertical sections (thin slices) of the sample \cite{Julia,Jacobs,neu}.

(d) X-Ray Tomography \cite{TOM}.

(e) Scanning tunneling microscopy (STM) \cite{STM,add2}.

To cover all relevant length scales one usually needs to combine several methods. Optical methods are limited
by the wavelength of the light and the shortest length scales that can be probed are usually of order $\sim 1 \ {\rm \mu m}$.
The (maximum) resolution of stylus measurements is determined by the radius of curvature $R$ of the probe tip 
and the (minimum) curvature radius $R^*$ of the valleys of the 
surface roughness profile (if $R>R^*$ the tip cannot penetrate into the cavity) \cite{pen1,pen2}. 
For engineering stylus instrument 
the tip curvature is typically $R \sim 1 \ {\rm \mu m}$ while for AFM very sharp tips with $R\sim 1 \ {\rm nm}$
are often used, and sometimes atomic structures (at nanometer length scales) can be resolved in AFM measurements.
However, surface roughness at the atomic scale is a somewhat ill-defined concept and the measured topography depends 
on the probe-tip substrate interaction potential and charge density profiles and involves quantum mechanical considerations.
However, there is really no limitation since the atomic scale ``roughness'' (e.g., atomic steps or adatoms) can anyhow not
be directly included in contact mechanics models (which are usually based on continuum mechanics), 
but require an atomistic approach such as Molecular Dynamics, so a ``multi-physics'' approach may be needed to fully understand a phenomena
depending on the surface roughness. Thus the shortest wavelength for which the power spectrum
is needed is of order $\sim 1 \ {\rm nm}$. 

The maximum lateral size of the studied surface area using AFM (and STM) is for most instruments of 
order $\sim 100 \ {\rm \mu m}$ so if longer wavelength roughness matters AFM must be combined with other methods.
Recently the use of optical methods has proliferated because of their simplicity and speed. These methods produce
nice looking pictures but we will show that the topography is not correctly reproduced in most cases and we
recommend against using optical methods for quantitative topography studies.

In this study, we will discuss the usefulness of different experimental methods on ``smooth'' and ``rough''
surfaces of elastically stiff and optically non-transparent solids. We will show how the 
2D power spectra, which enter contact mechanics theories, can be obtained from the 1D power spectra.
We find that in many cases optical methods fail while engineering stylus and AFM measurements usually give very good
results.

\vskip 0.3cm
{\bf 2 Surface roughness correlation functions}

Consider a surface with the height profile $z=h({\bf x})$ where ${\bf x}=(x,y)$ is a point
in the surface plane. We assume no overhangs so that for each point ${\bf x}$ there is only one height
coordinate $h({\bf x})$. We assume that the surface is nominally flat and that we choose the origin of the $z$-coordinate 
so that $\langle h({\bf x})\rangle = 0$, where $\langle .. \rangle$ stands for ensemble averaging.
In the most general case, to fully characterize the statistical properties of the surface, one needs to determine an infinite number of correlation functions:
$$\langle h({\bf x})h({\bf x'})\rangle , \ \ \ \langle h({\bf x})h({\bf x'})h({\bf x''})\rangle, \ \ ... $$

For randomly rough surfaces all the statistical properties of the surface are already contained in the
lowest-order correlation function:
$$C({\bf x},{\bf x'})=\langle h({\bf x})h({\bf x'})\rangle \eqno(1)$$
In this case, all correlation functions with an odd number of height coordinates $h({\bf x})$ vanish, and those with an even number of height coordinates can be written as products of the pair correlations function (1) (see Appendix A in Ref. \cite{Rev}).

If the statistical properties of the surface are translational invariant, the
correlation function (1) only depends on the 
coordinate difference ${\bf x}-{\bf x'}$, and no information will be lost if we
put ${\bf x}'={\bf 0}$ in (1). For such systems, it is more useful to study the correlation function (1)
in wavevector space ${\bf q}=(q_x,q_y)$ with the 2D surface roughness power spectrum defined as:
$$C_{\rm 2D}({\bf q})  = {1\over (2\pi )^2} \int d^2 x \ \langle h({\bf x}) h({\bf 0})
\rangle e^{i{\bf q}\cdot {\bf x}} \eqno(2)$$
For surfaces with roughness with isotropic properties, $C_{\rm 2D}({\bf q})$ depends only on the magnitude $q=|{\bf q}|$
of the wave vector. 

In calculating power spectra $C_{\rm 2D}({\bf q})$ 
from measured topography data, instead of using (2) it is more convenient
to use the Fast Fourier Transform Method to first calculate the height in wavevector space:
$$h({\bf q}) = {1\over (2\pi )^2} \int_{A_0} d^2x \ h({\bf x}) e^{-i{\bf q}\cdot {\bf x}} $$
from which one can obtain \cite{Rev}
$$C_{\rm 2D}({\bf q}) = {(2\pi )^2 \over A_0} |h({\bf q})|^2 $$
where $A_0=L_0^2$ is the studied surface area.

Randomly rough surfaces can be obtained by adding plane waves with random phases: 
$$h({\bf x}) = \sum_{\bf q} B_{\rm q} e^{i {\bf q}\cdot {\bf x}+i\phi_{\rm q}}\eqno(3)$$
where $\phi_{\rm q}$ are random numbers uniformly distributed between $0$ and $2 \pi$.
The parameter $B_{\rm q}$ in (3) can be written as $B_{\rm q} = (2 \pi /L) [C_{\rm 2D}({\bf q})]^{1/2}$.

The power spectrum (2) is of course also defined for surfaces with non-random roughness and can be used to express many physical quantities. For example, when an elastic solid (with Young's modulus $E$ and Poisson ratio $\nu$) is squeezed into complete contact with a rough flat rigid surface, the elastic energy stored at the interface due to the deformations induced by the surface roughness is given by:
$$U_{\rm el} ={1\over 2} E^* A_0 \int d^2q \ q C_{\rm 2D}({\bf q})\eqno(4)$$
where $A_0$ is the nominal surface area and $E^* = E/(1-\nu^2)$ is the effective Young's modulus, and we have assumed no interfacial friction. Other important quantities determined by $C_{\rm 2D}({\bf q})$ are the rms-roughness
$h_{\rm rms}$ and the rms slope $\xi$:
$$h_{\rm rms}^2 = \langle h^2 \rangle = \int d^2q \ C_{\rm 2D}({\bf q})\eqno(5)$$
$$\xi^2 = \langle (\nabla h)^2 \rangle = \int d^2q \ q^2 C_{\rm 2D}({\bf q})\eqno(6)$$
The parameters $h_{\rm rms}$ and $\xi$ are usually denoted $S_{\rm q}$ and $S_{\rm dq}$.

A randomly rough surface has a Gaussian height probability distribution:
$$P_h = {1\over (2\pi)^{1/2} h_{\rm rms}} e^{-(h/h_{\rm rms})^2/2} ,\eqno(7)$$
This equation implies that for an infinite system, there will be arbitrary high surface points. However, for any finite system, the probability of finding very high surface points is small. Non-random surfaces usually have non-Gaussian height probabilities, and for such surfaces skewness and kurtosis are very useful quantities that are defined as:
$$S_{\rm sk} = {\langle h^3 \rangle \over \langle h^2 \rangle^{2/3}}, \ \ \ \ S_{\rm ku} = 
{\langle h^4 \rangle \over \langle h^2 \rangle^2} \eqno(8)$$
where
$$\langle h^n \rangle = \int_{-\infty}^\infty dh \ h^n P_h$$
For randomly rough surfaces ($P_h$ is Gaussian), $S_{\rm sk}=0$ and $S_{\rm ku}=3$.

\vskip 0.3cm
{\bf 3 Surface roughness power spectra: theory}

The 2D power spectrum can be calculated from the height coordinate of a surface $z=h({\bf x})$ with ${\bf x} = (x,y)$, which is assumed given in $N\times N$ data points 
(typically $N=512$ or $1024$). The studied surface area may be part of the surface of a curved body but we assume that the macroscopic curvature is removed so that $\langle h \rangle = 0$. As discussed above, the roughness profile $z=h({\bf x})$ of a randomly rough surface can be written as a sum of plane waves $e^{i{\bf q}\cdot {\bf x}}$
with different wave vectors ${\bf q}$. The wavenumber $q=|{\bf q}| = 2 \pi /\lambda$ where $\lambda$ is the wavelength of one roughness component.

The most important property of a rough surface is its 2D power spectrum (2). Assuming that the surface has isotropic statistical properties,
$C_{\rm 2D}({\bf q})$ depends only on the magnitude $q= |{\bf q}|$ of the wave vector. A self-affine fractal surface has a power spectrum $C_{\rm 2D}(q)\sim q^{-2(1+H)}$ (where $H$ is the Hurst exponent related to the fractal dimension $D_{\rm f} = 3-H$), which is a straight line with the slope $-2(1+H)$ when plotted on a log-log scale. Most solids have surface roughness with the Hurst exponent $0.7 < H <1$ (see Ref. \cite{Per}).

For a one-dimensional (1D) line scan $z=h(x)$, one can calculate  the 1D power spectrum defined by:
$$C_{\rm 1D} (q) = {1\over 2 \pi} \int_{-\infty}^\infty dx \ \langle h(x) h(0) \rangle e^{i q x}\eqno(9)$$
Using (2) and (9) it is easy to show that the mean square (ms) roughness 
$$h^2_{\rm rms} = 2 \pi \int_{0}^\infty dk \ k C_{\rm 2D}(k) = 2 \int_{0}^\infty dk \ C_{\rm 1D}(k)\eqno(10) $$
If we assume the surface has self-affine fractal roughness, the power spectrum can be written as
$$C_{\rm 2D}(q) = C_0 q^{-2-2H}$$
which is defined for $q_0 < q < q_1$, and zero otherwise, then using (10) we get 
$$h_{\rm rms}^2 = 2 \pi \int_{q_0}^{q_1} dk \ C_0 k^{-1-2H} \approx {\pi \over H} C_0 q_0^{-2H}$$
where we have assumed $q_1/q_0 >>1$ and $H>0$, thus
$$C_{\rm 2D}={H \over \pi } h_{\rm rms}^2 {1\over q_0^2} \left ({q\over q_0}\right )^{-2-2H}\eqno(11)$$
In a similar way using (9) and (10) we get
$$C_{\rm 1D}=H h_{\rm rms}^2 {1\over q_0} \left ({q\over q_0}\right )^{-1-2H}\eqno(12)$$
Note that $C_{\rm 1D}=\pi q C_{\rm 2D}$ but this relation only holds exactly when $C_{\rm 2D}$ is a power law in the wavenumber $q$ (see below).

For surfaces with isotropic roughness the 2D power spectrum $C_{\rm 2D} (q)$ can be obtained directly
from $C_{\rm 1D} (q)$ (see \cite{Carbone}):
$$C_{\rm 2D}(q) = {1\over \pi}\int_q^{\infty} dk \ {[-C'_{\rm 1D}(k)]\over \left (k^2-q^2\right )^{1/2}}\eqno(13)$$
This relation for calculating the 2D power spectrum from the 1D power spectrum is very useful.
The related relation on how to obtain the 1D power spectrum from the 2D power spectrum is less useful \cite{add1}:
$$C_{\rm 1D}(q) = 2 \int_q^{\infty} dk \ {k C_{\rm 2D}(k)\over \left (k^2-q^2\right )^{1/2}}\eqno(14)$$
If $C_{\rm 1D}(q)$ is obtained from line scan measurements it is known only numerically. And since it is in general a noisy function of the wavenumber $q$,
the derivative $C'_{\rm 1D}(q)$ will typically fluctuate wildly with $q$ making the relation (13) not very useful in many practical applications unless $C_{\rm 1D}(q)$ can be fitted by a smooth function. 

Equations (13) and (14) are exact and must hence obey the ``sum rule'' (10). To demonstrate that, taking (13) as an example, we calculate
$$I=2 \pi \int_0^\infty dq \ q C_{\rm 2D} (q)$$
$$ = 2 \pi \int_0^\infty dq \ q {1 \over \pi} \int_q^{\infty} 
dk \ {[-C'_{\rm 1D}(k)]\over \left (k^2-q^2\right )^{1/2}}$$ 
Writing $k=q x$ this gives
$$I=2 \int_0^\infty dq \ q \int_1^{\infty} 
dx \ {[-C'_{\rm 1D}(qx)]\over \left (x^2-1\right )^{1/2}}$$
$$ = 2 \int_1^{\infty} dx \ {1\over \left (x^2-1\right )^{1/2}}  
\int_0^\infty dq \ q [-C'_{\rm 1D}(qx)]$$
Defining $qx=y$ this gives
$$I=2 \int_1^{\infty} dx \ {1\over x^2 \left (x^2-1\right )^{1/2}} \int_0^\infty dy \ y [-C'_{\rm 1D}(y)]$$
Using partial integration and assuming that $y C_{\rm 1D}(y)$ vanishes for $y=0$ and $y=\infty$ we get
$$I = 2 \int_0^\infty dy  \ C_{\rm 1D}(y) = h_{\rm rms}^2$$
where we have used that
$$\int_1^{\infty} dx \ {1\over x^2 \left (x^2-1\right )^{1/2}} =1$$
an equation that is easy to prove by changing the integration variable using $x=1/{\rm cos}\phi$.
Similarly, one can show that (14) satisfies the sum rule (10).

There are two other way to calculate $C_{\rm 2D}(q)$ approximately from $C_{\rm 1D}(q)$.
For a self-affine fractal surface (without a roll-off) fitting $C_{\rm 1D}(q)$ with (12) 
gives the Hurst exponent $H$ and the rms roughness $h_{\rm rms}$
and knowing these parameters from (11) we can calculate $C_{\rm 2D}(q)$ (see also Appendix A).

\begin{figure}[!ht]
        \includegraphics[width=0.45\textwidth]{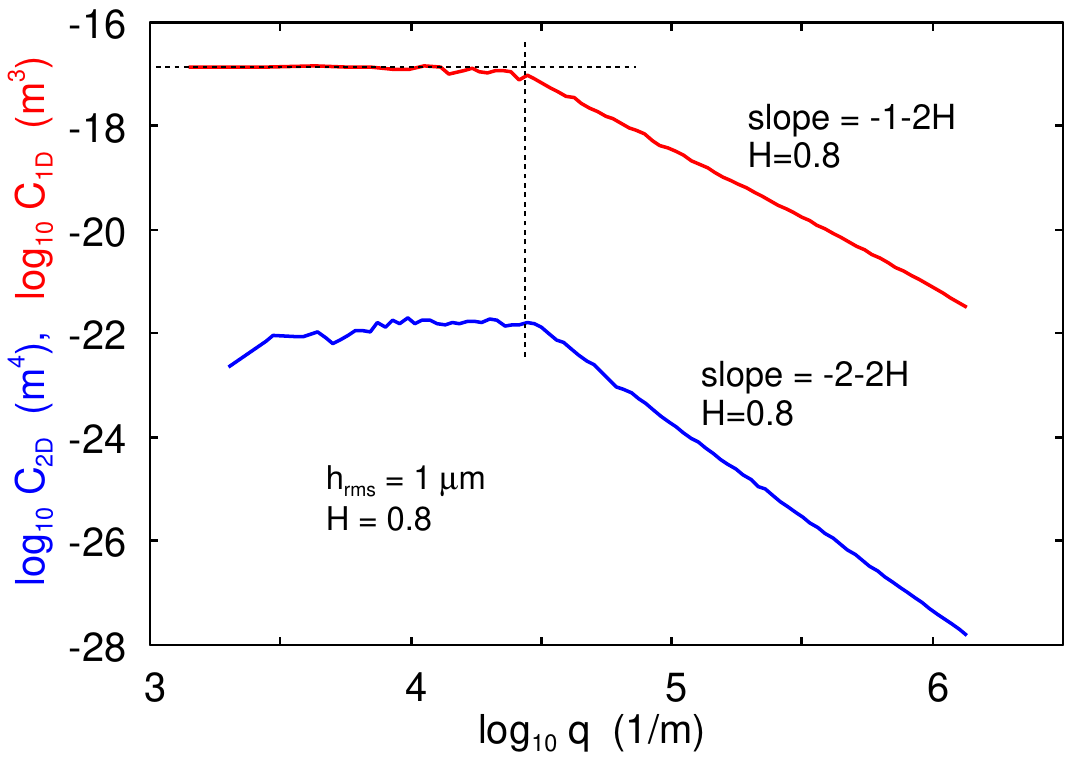}
\caption{\label{1logq.2logC.red1D.blue2D.eps}
The 1D (red line) and 2D (blue line) power spectra calculated from a mathematically generated randomly rough surface with
the rms roughness $h_{\rm rms} = 1 \ {\rm \mu m}$ and Hurst exponent $H=0.8$. The small, large and
roll-off wavenumbers are $q_0=1\times 10^3 \ {\rm m}^{-1}$, $q_1=2048 \times 10^3 \ {\rm m}^{-1}$
and $q_{\rm r}=30 \times 10^3 \ {\rm m}^{-1}$, respectively.
}
\end{figure}

A second way is as follows. First note that the mean square (ms) roughness amplitude can be obtained from $C_{\rm 1D}(q)$ and $C_{\rm 2D}(q)$ via (10).
If the topography is measured with an instrument with finite resolution then only wavenumbers below
some value $q = 2 \pi /\lambda$ can be resolved. In that case, one would only observe an apparent ms roughness
$$h^2_{\rm rms} \approx 2 \pi \int_{0}^{q} dk \ k C_{\rm 2D}(k) \approx 2 \int_{0}^{q} dk \ C_{\rm 1D}(k) $$
In this equation, $q$ can be considered as a parameter (depending on the instrument resolution) so we can 
take the derivative with respect to $q$ to get
$$2 \pi q C_{\rm 2D}(q) \approx 2 C_{\rm 1D}(q)$$
or
$$C_{\rm 2D}(q) \approx {1\over \pi q} C_{\rm 1D}(q)\eqno(15)$$
which is the same result as found previously by comparing (11) and (12).
However, this relation is in general only approximate and the exact relation is given by (13).
Note in particular if $C_{\rm 1D}(q)$ has a roll-off region for $q<q_{\rm r}$, where $C_{\rm 1D}$ is approximately
constant, then from (15) we get in the roll-off region $C_{\rm 2D} \sim 1/q$ while the exact result (13) for $q<<q_{\rm r}$ gives
$$C_{\rm 2D}(q) \approx {1\over \pi}\int_{q_{\rm r}}^{q_1} dk \ {1\over k} [-C'_{\rm 1D}(k)]$$
which is constant. Thus if (15) is used for a self-affine fractal surface with a roll-off, one should replace
the $\sim 1/q$ region with a constant determined by $C_{\rm 2D}$ at the onset $q=q_{\rm r}$
of the roll-off region (here $q_{\rm r}$ refers to the largest wavenumber $q$ of the
$\sim 1/q$ region). In addition the magnitude of $C_{\rm 2D}(q)$ given by (15)
in the self-affine fractal region must be corrected (scaled)
by a factor $\approx (1+3H)^{1/2}$ which depends on the Hurst exponent $H$ (see Appendix A). 
This correction factor is needed in order for the sum-rule (10) to be satisfied: decreasing $C_{\rm 2D}(q)$ in the roll-off region
imply we must increase $C_{\rm 2D}(q)$ in the self-affine fractal region in order for the sum-rule to be obeyed.
Note that the onset of the roll-off region occurs
at the same wavenumber $q_{\rm r}$ for $C_{\rm 1D}(q)$ and $C_{\rm 2D}(q)$ (see vertical dashed line in
Fig. \ref{1logq.2logC.red1D.blue2D.eps}). In most practical applications, the detailed form of the power spectrum
in the roll-off region is not very important (but the fact that a roll-off region exists for $q<q_{\rm r}$ is very important). 

\begin{figure}[!ht]
        \includegraphics[width=0.45\textwidth]{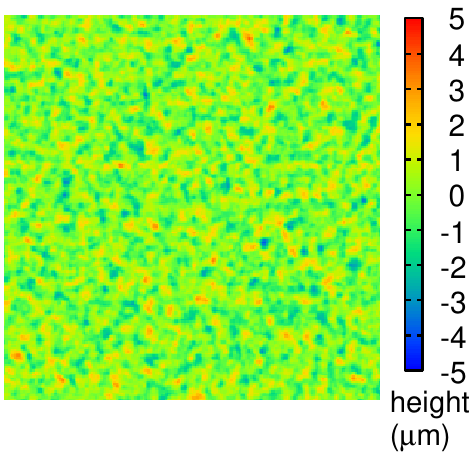}
\caption{\label{1x.2y.3h.plot.mathematical.eps}
A mathematically generated randomly rough surface with
the rms roughness $h_{\rm rms} = 1 \ {\rm \mu m}$ and Hurst exponent $H=0.8$. The small, large and
roll-off wavenumbers are $q_0=1\times 10^3 \ {\rm m}^{-1}$, $q_1=2048 \times 10^3 \ {\rm m}^{-1}$
and $q_{\rm r}=30 \times 10^3 \ {\rm m}^{-1}$, respectively.
}
\end{figure}

\begin{figure}[!ht]
        \includegraphics[width=0.45\textwidth]{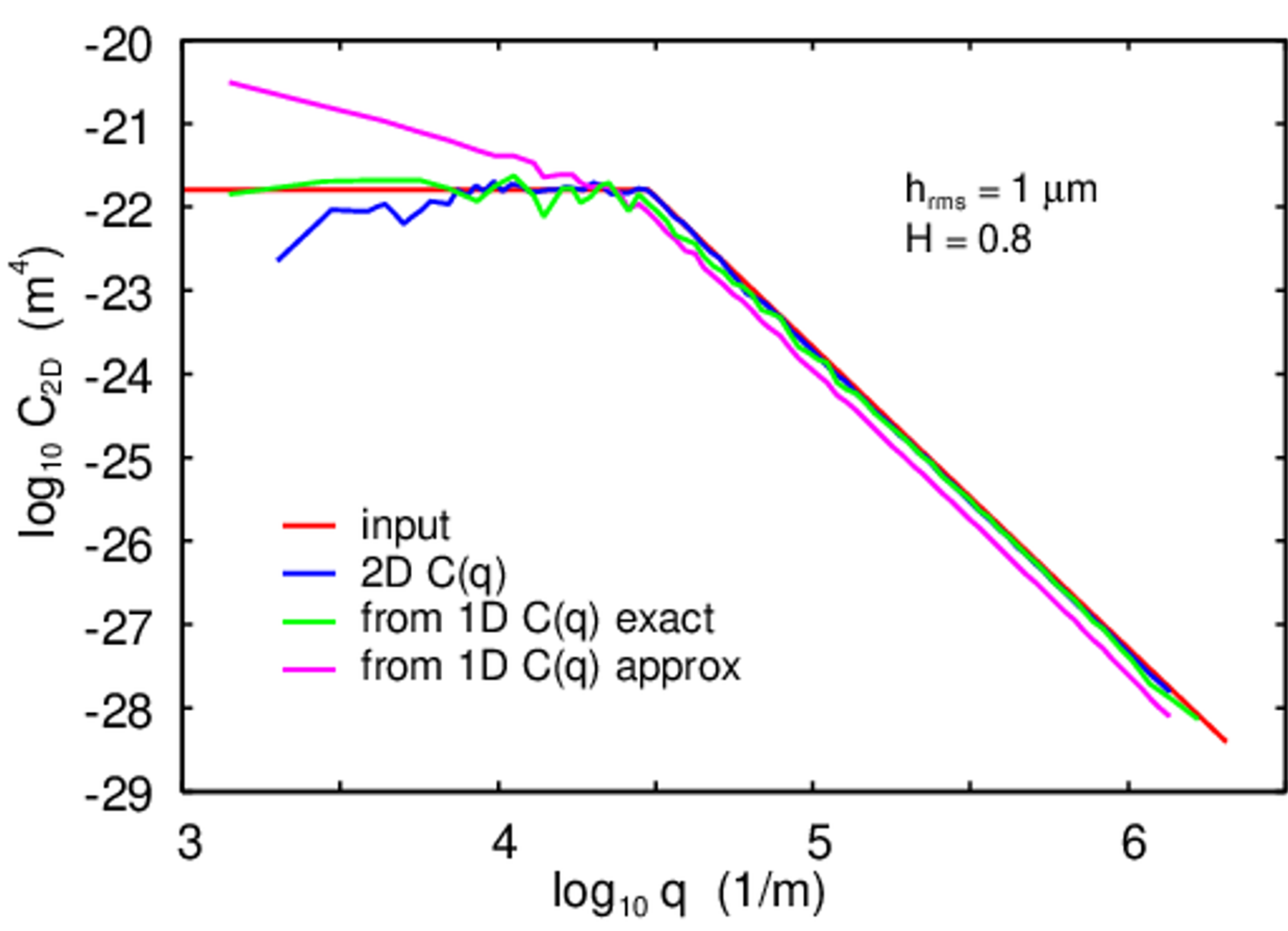}
\caption{\label{1logq.2logC.redInput.blue2D.greenExact1D.pink.2TimesAsymetry.eps}
The input power spectrum is used to generate the surface topography of a randomly rough surface (red line),
and the power spectra are calculated from the rough surface using the 2D power spectrum equation (blue line),
using (13) with the 1D power spectrum calculated from the rough surface (red line in Fig. \ref{1logq.2logC.red1D.blue2D.eps})
(green line) and using the approximate equation (15) (pink line).
}
\end{figure}

To illustrate the discussion above we have calculated the power spectrum for a randomly rough
surface obtained by adding plane waves with random phases (see (3) and Appendix A in Ref. \cite{Rev}).
Fig. \ref{1x.2y.3h.plot.mathematical.eps} shows a mathematically generated randomly rough surface with
the rms roughness $h_{\rm rms} = 1 \ {\rm \mu m}$ and Hurst exponent $H=0.8$. The small, large and
roll-off wavenumbers are $q_0=1\times 10^3 \ {\rm m}^{-1}$, $q_1=2048 \times 10^3 \ {\rm m}^{-1}$
and $q_{\rm r}=30 \times 10^3 \ {\rm m}^{-1}$, respectively. The surface consists of $2048 \times 2048$
height data points.

Fig. \ref{1logq.2logC.red1D.blue2D.eps}
shows the 1D (red line) and 2D (blue line) power spectra calculated from 
the height profile $z=h(x,y)$ of the surface shown in Fig. \ref{1x.2y.3h.plot.mathematical.eps}.
The 1D power spectrum is the average of $C_{\rm 1D}$ over all 2048 $z=h(x)$ ($y$ constant) lines contained in the height profile $z=h(x,y)$. 

The red line in Fig. \ref{1logq.2logC.redInput.blue2D.greenExact1D.pink.2TimesAsymetry.eps}
shows the input power spectrum used to generate the surface 
shown in Fig. \ref{1x.2y.3h.plot.mathematical.eps}. The blue line is the
power spectra calculated from the rough surface using the 2D power 
spectrum equation (2). The green line is the 2D power spectrum
obtained from the 1D power spectrum (given by the red line in 
Fig. \ref{1logq.2logC.red1D.blue2D.eps}) using the (exact) equation (13),
while the pink line is calculated using the approximate equation (15).
Note that the approximate formula (15) gives a linear dependency of $1/q$ in the roll-off region in contrast to the nearly constant value resulting from the exact equation (13). Additionally, in the self-affine fractal region, the power spectrum as described by (15) is underestimated by a factor of $\approx (1+3H)^{1/2}$. For $H=0.8$, this factor is approximately $1.84$ (see Appendix A).

\vskip 0.3cm
{\bf 4 Surface roughness power spectra: experiment}

Using optical and stylus measurements, we have studied the surface topography on two types of samples supplied by the Surface Topography Challenge \cite{Chal}. We refer to these two types as the ``rough'' and ``smooth'' surfaces. In this study, six smooth samples: B54, B55, C13, C14, A71, A72 and five rough samples: R54, P85, P86, P63, P64 were used. The same samples were prepared on one single wafer and subsequently cut into $1 \ {\rm cm} \times 1  \ {\rm cm}$ pieces, so theoretically they should have similar surface characteristics.

Both smooth and rough samples have coatings made from chromium nitride (CrN). The smooth surface has CrN deposited on a prime-grade polished silicon wafer, while the rough surface has CrN deposited on the rough ``backside'' of a single-side-polished silicon wafer, which has been subsequently etched with isotropic reactive ion. CrN was chosen because it is a wear-and-corrosion-resistant coating which is widely used in automotive components, cutting tools, and die-casting. CrN is typically deposited via physical vapor deposition (PVD), and the present deposition uses a magnetron sputtering technique. The silicon substrates were chosen due to their extreme reproducibility in fabrication. The two substrates are intended to produce a ``smooth surface'' that is representative of materials used in the semiconductor industry, and a ``rough surface'' that has larger topographic variation, as is common in other industrial contexts.

\begin{figure}[!ht]
        \includegraphics[width=0.45\textwidth]{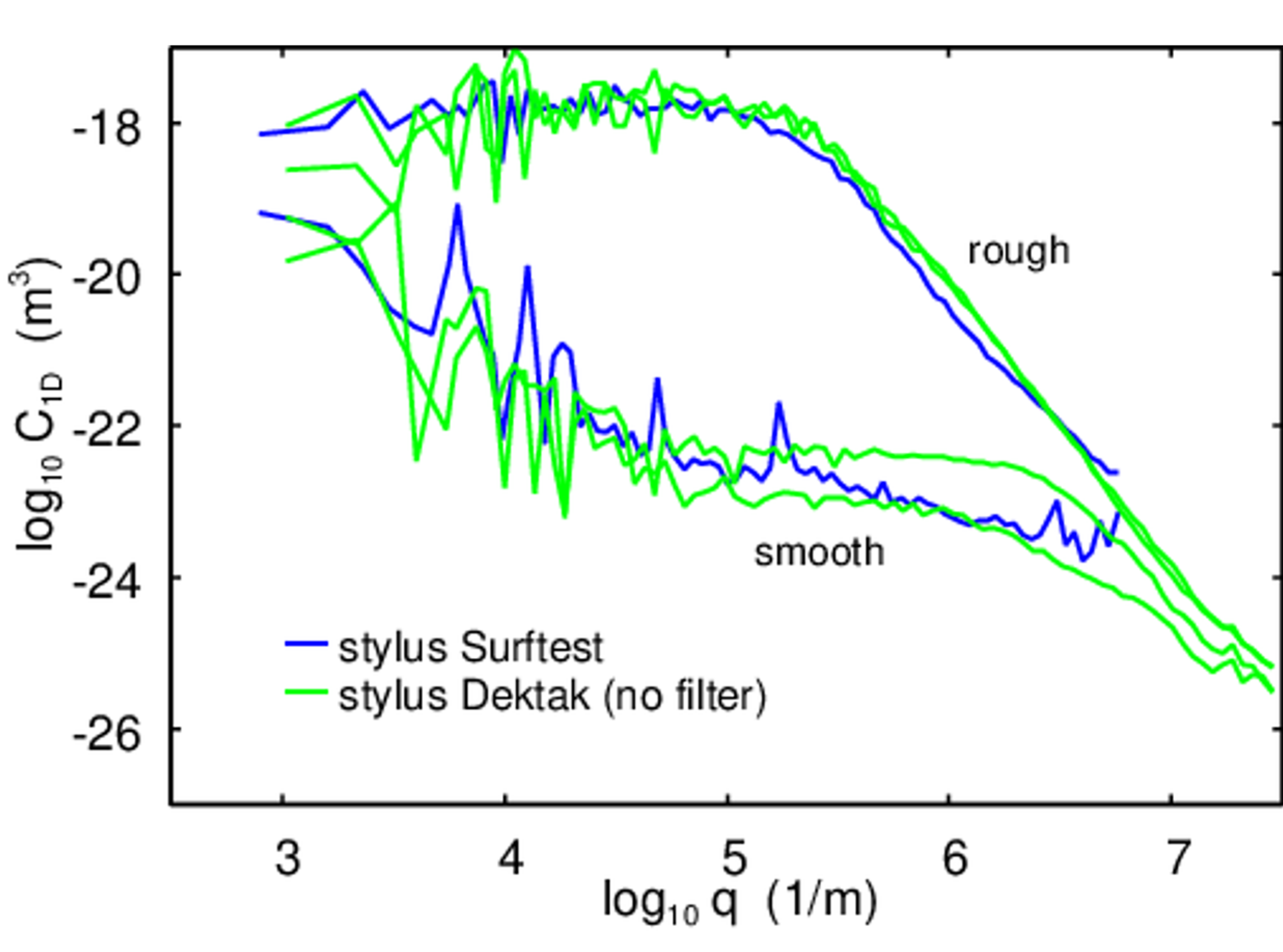}
\caption{\label{1logq.2logC.Julia.and.BD.stulus.repeat.eps}
The 1D surface roughness power spectra of the smooth and rough surfaces calculated from topography data obtained using two different
engineering stylus instruments. The blue lines are the stylus data shown in Fig. \ref{1logq.2logC1D.AFMandJulia.eps}
while the green lines are using another (Dektak) stylus instrument.
The two green lines for the smooth and rough surfaces were obtained on different sample surfaces (samples B55 and C13 for the smooth surface and R54 and P86 for the rough surface).
}
\end{figure}

\begin{figure}[!ht]
        \includegraphics[width=0.45\textwidth]{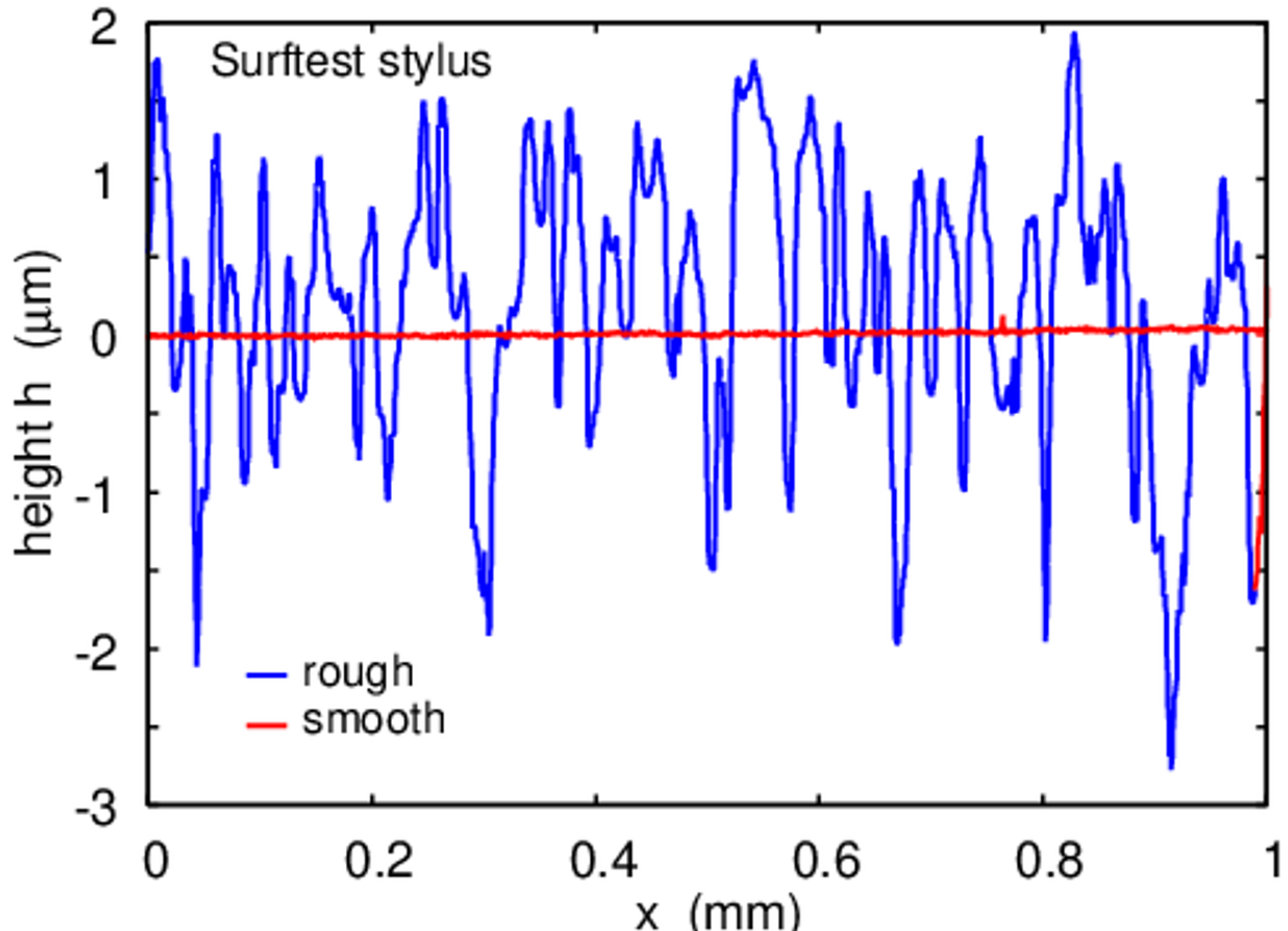}
\caption{\label{1x.2h.Julia.eps}
The height $h(x)$ as a function of $x$ for the smooth (red) and rough (blue) surfaces as obtained using
the Surftest stylus instrument.
}
\end{figure}

\begin{figure}[!ht]
        \includegraphics[width=0.45\textwidth]{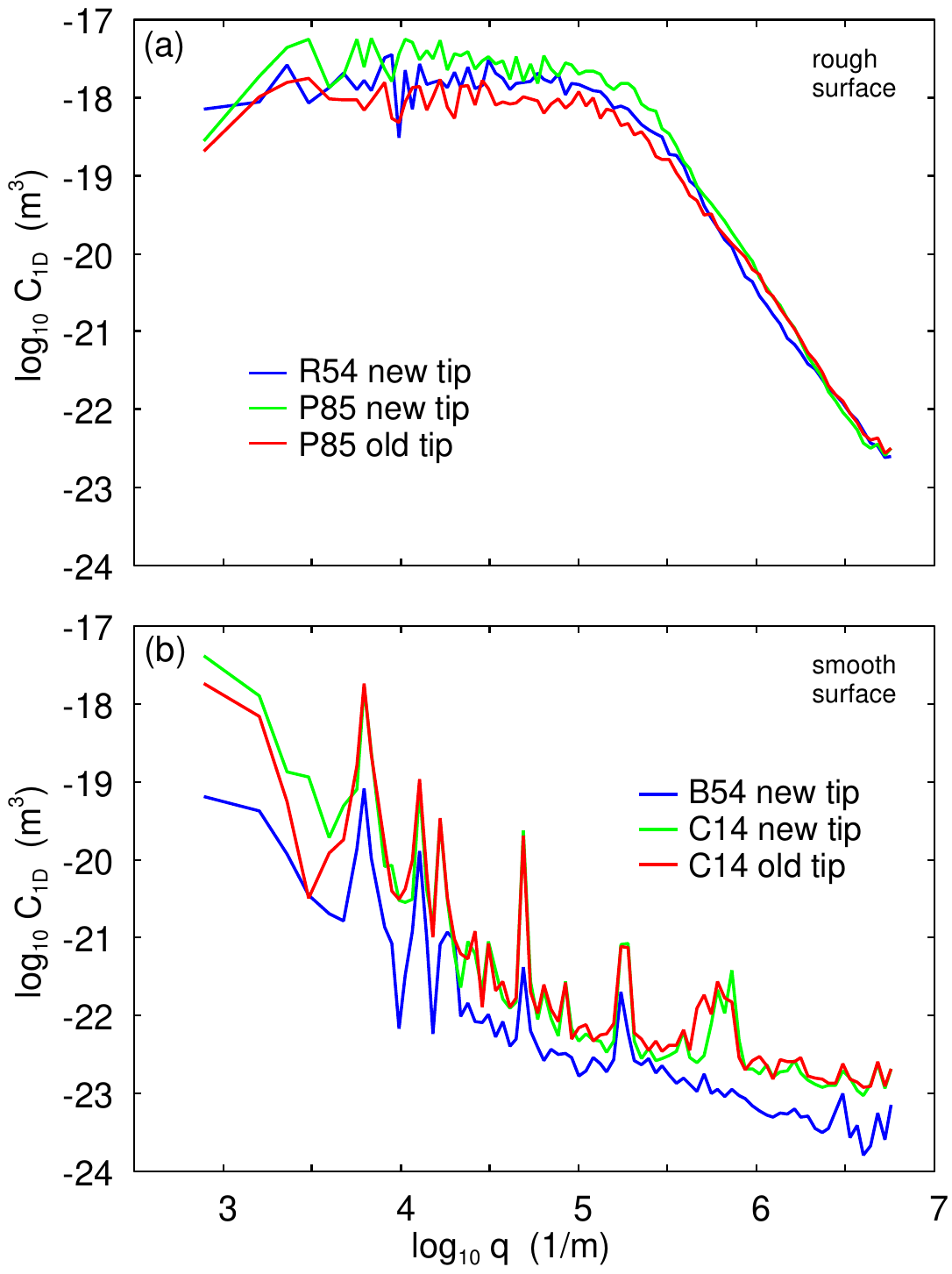}
\caption{\label{smooth.rough.JULIA.eps}
The power spectrum calculated from the Mitutoyo topography data
for the samples R54 (rough) and B54 (smooth) (blue lines, from Fig. \ref{1logq.2logC.Julia.and.BD.stulus.repeat.eps})
and for the surfaces P85 (rough) and C14 (smooth) using an old (red lines) and new (green lines) diamond tip. The old tip was used
for several years (for several 100 topography measurements) before performing the measurements reported here.
}
\end{figure}

\vskip 0.1cm
{\bf Engineering stylus}

The topography measurements were performed using two stylus profilometers:

(a) A Mitutoyo Portable Surface Roughness Measurement Surftest SJ-410 equipped with a diamond tip having a radius of curvature $R=1 \ {\rm \mu m}$, and with the tip-substrate repulsive force $F_N = 0.75 \ {\rm mN}$. The step length (pixel) is $0.5 \ {\rm \mu m}$, the scan length $L=8 \ {\rm mm}$ and the tip speed  $v=50 \ {\rm \mu m/s}$. The power spectra shown in Fig. \ref{1logq.2logC.Julia.and.BD.stulus.repeat.eps} (blue lines) were obtained by averaging over three measurements on each surface.

(b) A Bruker Dektak XT, equipped with a diamond tip having a radius of curvature $R = 0.7 \ {\rm \mu m}$, and with the tip-substrate repulsive forces $F_{\rm N} = 1\times 10^{-5} \ {\rm N}$ and $2\times 10^{-5} \ {\rm N}$. The scan lengths were $L=3$ and $6 \ {\rm mm}$ with steps resolutions of $0.15$ and $0.10 \ {\rm \mu m}$, respectively. The tip speed varied between $33$ and $44 \ {\rm \mu m/s}$. The Bruker Dektak was mounted on a vibration isolation table.

The blue and green lines in 
Fig. \ref{1logq.2logC.Julia.and.BD.stulus.repeat.eps} are the 1D surface roughness
power spectra of the smooth and rough surfaces obtained with the Mitutoyo and
Bruker stylus instruments, respectively.
The two green lines for the smooth and rough surfaces
where obtained on different sample surfaces (samples B55 and C13 for the smooth surface and R54 and P86 for the rough surface).
In our original measurement with the Brucker stylus, a short wavelength filter was included which resulted
in large deviations between the two stylus measurements for large wavevectors (not shown).
Filters are used to make some quantities, such as the rms slope, more well-defined \cite{WeeDefined}. However, many physical quantities, such as the area of real contact, depend on the short wavelength roughness. Therefore, \emph{no filter should be used in calculating power spectra that are intended for theoretical calculations}.

Fig. \ref{1x.2h.Julia.eps}
shows the height $h(x)$ as a function of $x$ for the smooth (red) and rough (blue) surfaces as obtained using
the Surftest stylus instrument.

Fig. \ref{smooth.rough.JULIA.eps}
shows the power spectrum calculated from the Mitutoyo data
for the samples R54 (rough) and B54 (smooth) (blue lines, from Fig. \ref{1logq.2logC.Julia.and.BD.stulus.repeat.eps})
and for the surfaces P85 (rough) and C14 (smooth) using an 
old (red lines) and new (green lines) diamond tip. The old tip was used
for several years (for several hundred topography measurements) before performing the measurements reported here.  
The blue curve in Fig. \ref{smooth.rough.JULIA.eps} was measured on B54 (smooth wafer)  $\sim 1/2$ year
after the measurements of the red and green curves. For the C14 surface the old and the new tips give nearly the same power
spectra which indicates that wear (or contaminations) on the tips has a negligible influence on the measured topography. 


        \begin{figure}[!ht]
        \includegraphics[width=0.45\textwidth]{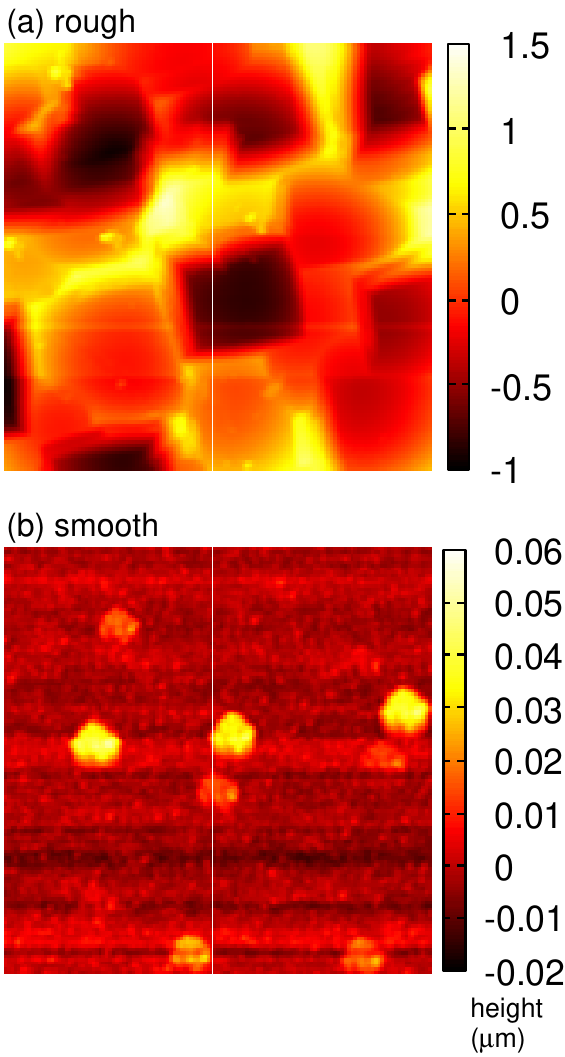}
\caption{\label{pic.RoughSmooth.eps}
AFM pictures of the rough and the smooth surfaces. The studied unit is $40 \times 40 \ {\rm \mu m}^2$ 
for the rough surface and  $5 \times 5 \ {\rm \mu m}^2$ for the smooth surface. 
The square units in (a)
have the linear size $\sim 10 \ {\rm \mu m}$.
}
\end{figure}

\begin{figure*}[!ht]
        \includegraphics[width=0.95\textwidth]{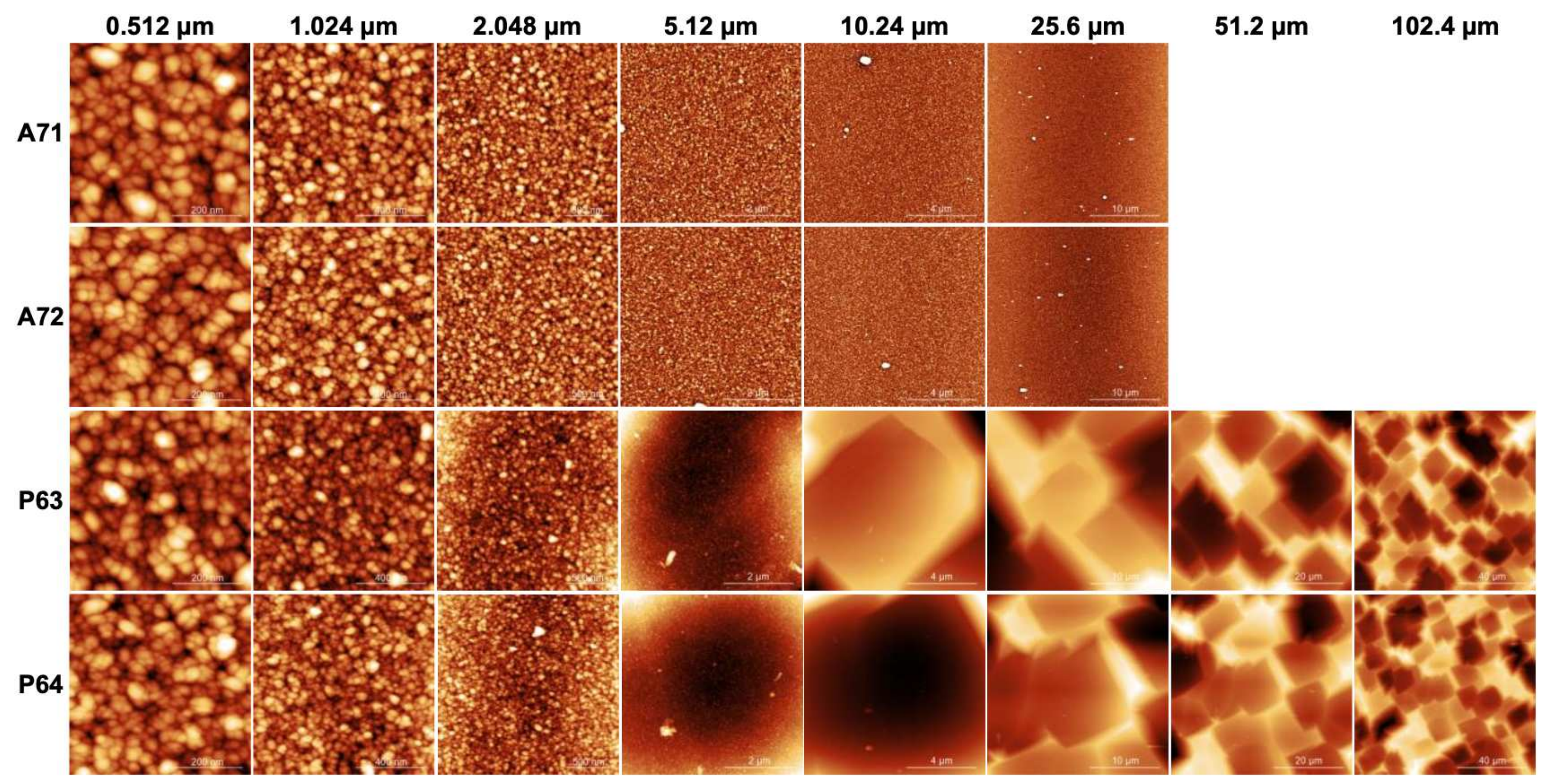}
\caption{\label{AFMpicAll.ps}
AFM topography images at different scan sizes for smooth (A71, A72) and rough (P63, P64) surfaces.
}
\end{figure*}

\vskip 0.1cm
{\bf Atomic Force Microscopy (AFM)}

We have used two AFM setups in the present study:

(a) A Bruker Dimension 3100 AFM in tapping mode (amplitude modulation) equipped with RSTESPA-300 probe having a $8 \ {\rm nm}$ tip radius. 
Two 2D scan pixel numbers were used: $512 \times 512$ and $1024 \times 1024$. The scanned
areas were as follows:  $5 \times 5 \ {\rm \mu m}^2$ for the smooth surfaces, and $40 \times 40 \ {\rm \mu m}^2$ 
for the rough surfaces. The reason for a larger scan length
of the rough surfaces was to have a good average over the observed ``plateau'' topography features. Fig. \ref{pic.RoughSmooth.eps} shows AFM topography images obtained with setup (a) for rough and smooth samples. 

(b) Veeco Multimode IIIA AFM in tapping mode. Three silicon ContAl-G cantilever probes (BudgetSensors, Bulgaria) were used in the measurements, one for sample A71, one for A72 and P63, and one for P64. The cantilevers have resonance frequencies of $\sim 14 \ {\rm kHz}$, the spring constants of $\sim 0.2 \ {\rm N/m}$, and tip apex radii of $\sim 10 \ {\rm nm}$. The scanning setpoints and the feedback parameters were manually optimized for each imaging. For the smoother A71 and A72 samples, multiple topographies were measured with scan sizes of 0.512, 1.024, 2.048, 5.12, 10.24, $25.6 \ {\rm \mu m}$. For the rougher P63 and P64 samples, two additional larger scan sizes of 51.2 and $102.4 \ {\rm \mu m}$ were also included. The scan rates used in the measurements are 3 lines/s for scan sizes of $0.512 \ {\rm \mu m}$, 2 lines/s for scan sizes of 1.024, $2.048  \ {\rm \mu m}$, and 1 line/s for others. All the measured topography images have data points of $512 \times 512$ pixels. Fig. \ref{AFMpicAll.ps} shows AFM topography images obtained with setup (b) for rough and smooth samples at different scan sizes.

\begin{figure}[!ht]
        \includegraphics[width=0.45\textwidth]{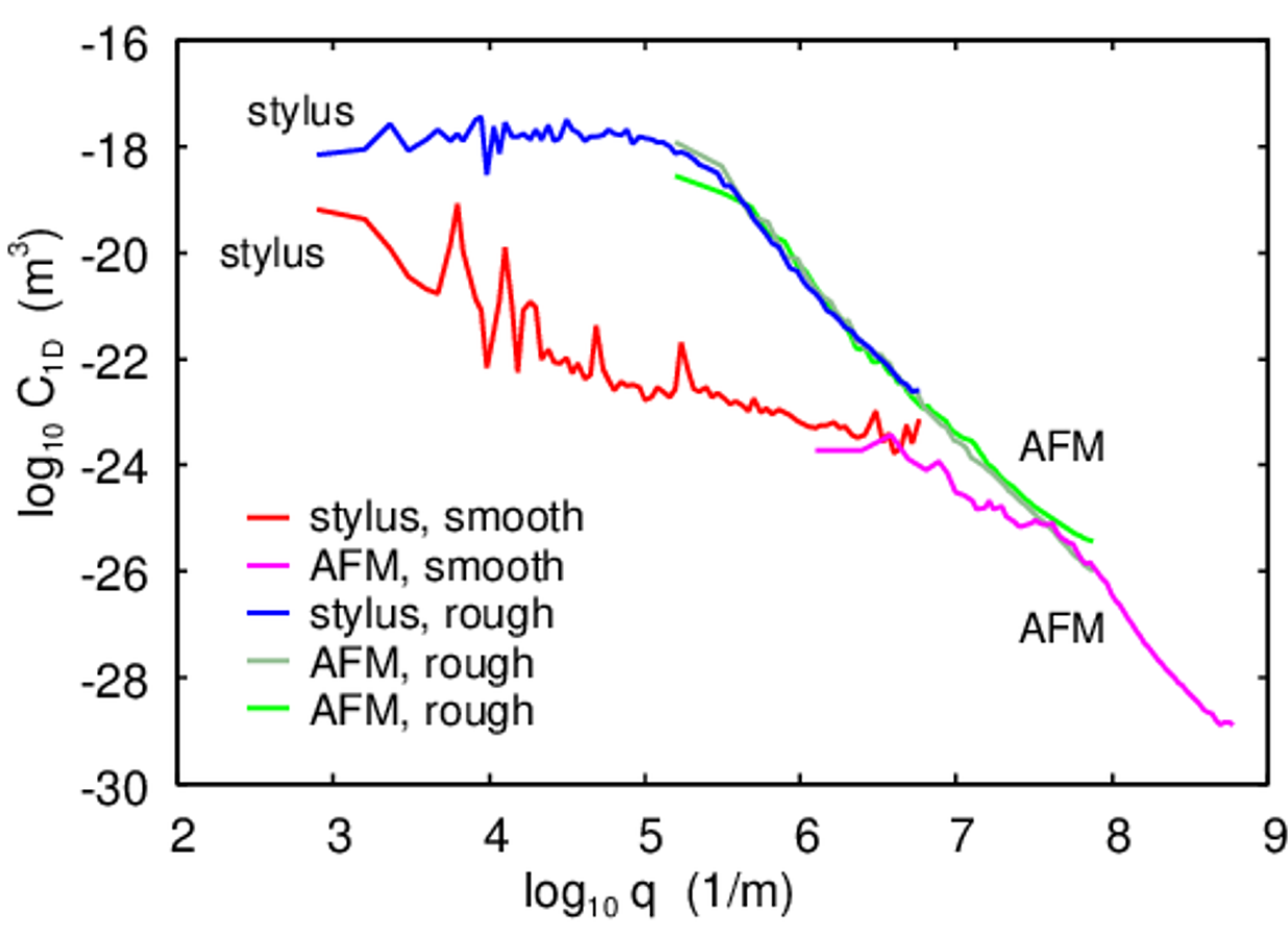}
\caption{\label{1logq.2logC1D.AFMandJulia.eps}
The 1D surface roughness power spectra of the smooth and rough surfaces calculated from topography data obtained using an
engineering stylus instrument (red and blue curves), and AFM setup (a). The stylus power spectrum was obtained by averaging
the power spectra obtained from 3 line scans each $8 \ {\rm mm}$ long. The AFM data consisted of $1024 \times 1024$ data points.
For the AFM data, the 1D power spectrum was calculated for each $z=h(x)$ line and averaged over all the 1024 lines.
}
\end{figure}

Fig. \ref{1logq.2logC1D.AFMandJulia.eps} shows the 1D power spectra calculated from the AFM topography data obtained with setup (a), for two measurements on the rough surface and one measurement on the smooth surface. In all cases, the topography was measured in $1024 \times 1024$ points. Note that the AFM data overlap the stylus measurements in the region of common wavenumber. This is a good test which shows that both measurement methods are accurate. Note also that the power spectra of the two surfaces for wavenumber $q > 3\times 10^7 \ {\rm m}^{-1}$ (or wavelength below $\lambda = 2\pi /q \approx 200 \ {\rm nm}$) appear to be nearly the same which may reflect some intrinsic roughness of the CrN coating independent of the substrate roughness. This is confirmed by AFM at higher resolution (Fig. \ref{AFMpicAll.ps}), which shows nearly the same topography for the smooth and rough surfaces at the highest resolution. Note that the short wavelength roughness of the substrate (with a wavelength much smaller than the thickness of the coating) may not show up at the surface of the coating film.

\begin{figure}[!ht]
        \includegraphics[width=0.45\textwidth]{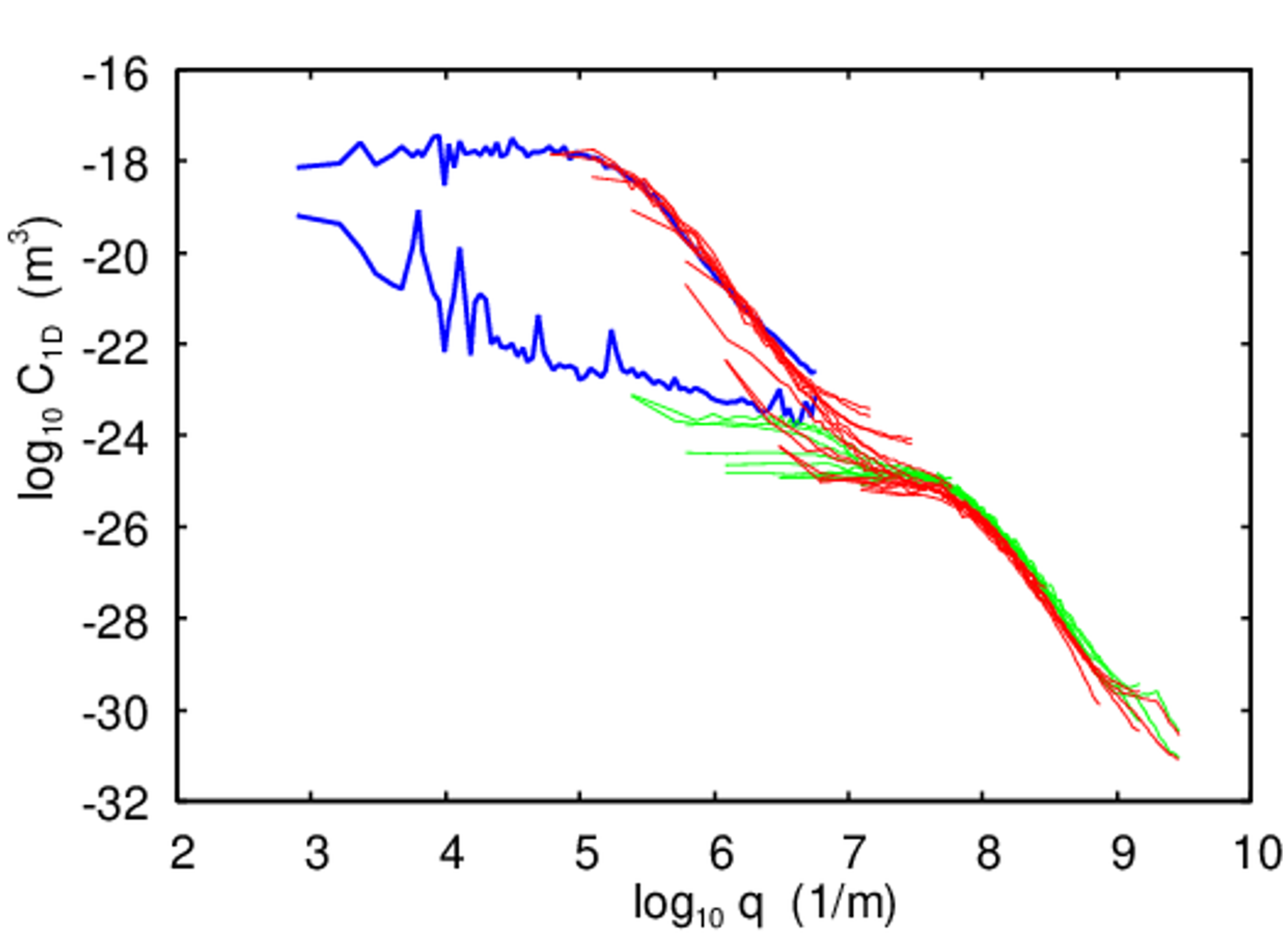}
\caption{\label{1logq.2logC.china2.eps}
The 1D surface roughness power spectra of the smooth and rough surfaces calculated from topography data obtained using an
engineering stylus instrument (blue curves), and AFM setup (b). The stylus power spectrum was obtained by averaging
the power spectra obtained from 3 line scans each $8 \ {\rm mm}$ long. The AFM data consisted of $512 \times 512$ data points.
For the AFM data, the 1D power spectrum was calculated for each $z=h(x)$ line and averaged over all the 512 lines.
}
\end{figure}

Fig. \ref{1logq.2logC.china2.eps} shows the 1D surface roughness power spectra of the smooth and rough surfaces 
calculated from topography data obtained using an
engineering stylus instrument (blue curves from Fig. \ref{1logq.2logC.Julia.and.BD.stulus.repeat.eps}), and the AFM setup (b). 
The stylus power spectrum was obtained by averaging
the power spectra obtained from 3 line scans each $8 \ {\rm mm}$ long.
For the AFM data the 1D power spectrum was calculated for each $z=h(x)$ 
line and averaged over all the 512 lines.

\vskip 0.1cm
{\bf Confocal Laser Scanning Profilometry (CLSP)} 

\begin{figure}[!ht]
        \includegraphics[width=0.45\textwidth]{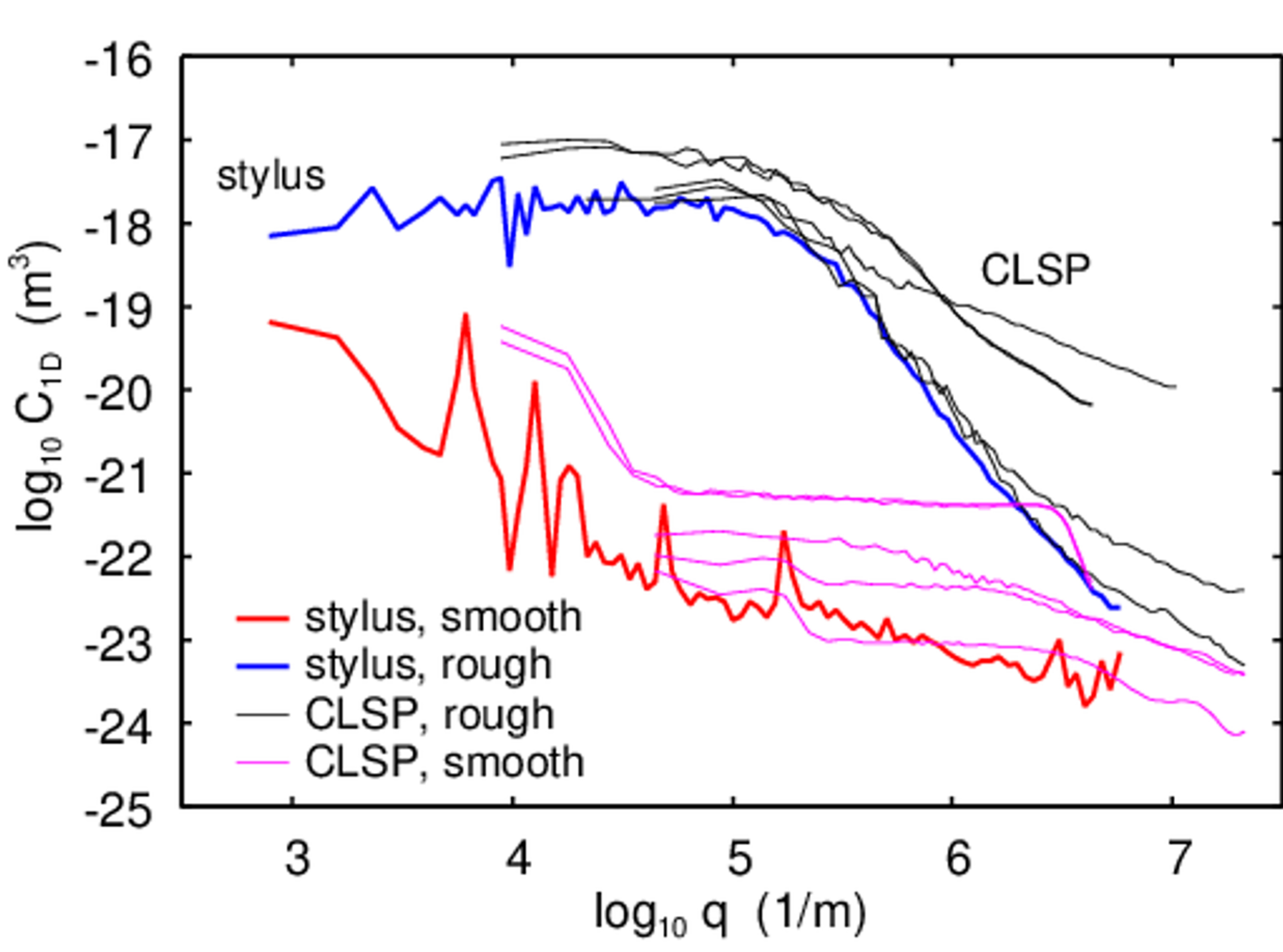}
\caption{\label{1logq.2logC.Julia.KEY1.1D.eps}
The 1D surface roughness power spectra of the smooth and rough surfaces calculated from topography data obtained using an
engineering stylus (red and blue curves), and 3D Confocal Laser Scanning Profilometry (CLSP). The CLSP 
was performed in $1024 \times 768$ data points.
}
\end{figure}

The laser profilometry was performed using two different
instruments: Keyence VK-X1050 and Keyence VK-3000.
Both instruments have magnifications of 10, 20, 50, and $100\times$. A red
$661 \ {\rm nm}$ laser beam was used and the pixel number was $768 \times 1024$.
Fig. \ref{1logq.2logC.Julia.KEY1.1D.eps} compares the $C_{\rm 1D}(q)$ results from the CLSP to the output from
engineering stylus for the rough and smooth surfaces.

\vskip 0.1cm
{\bf  Light Interferometry Profilometry (LIP)}

\begin{figure}[!ht]
        \includegraphics[width=0.45\textwidth]{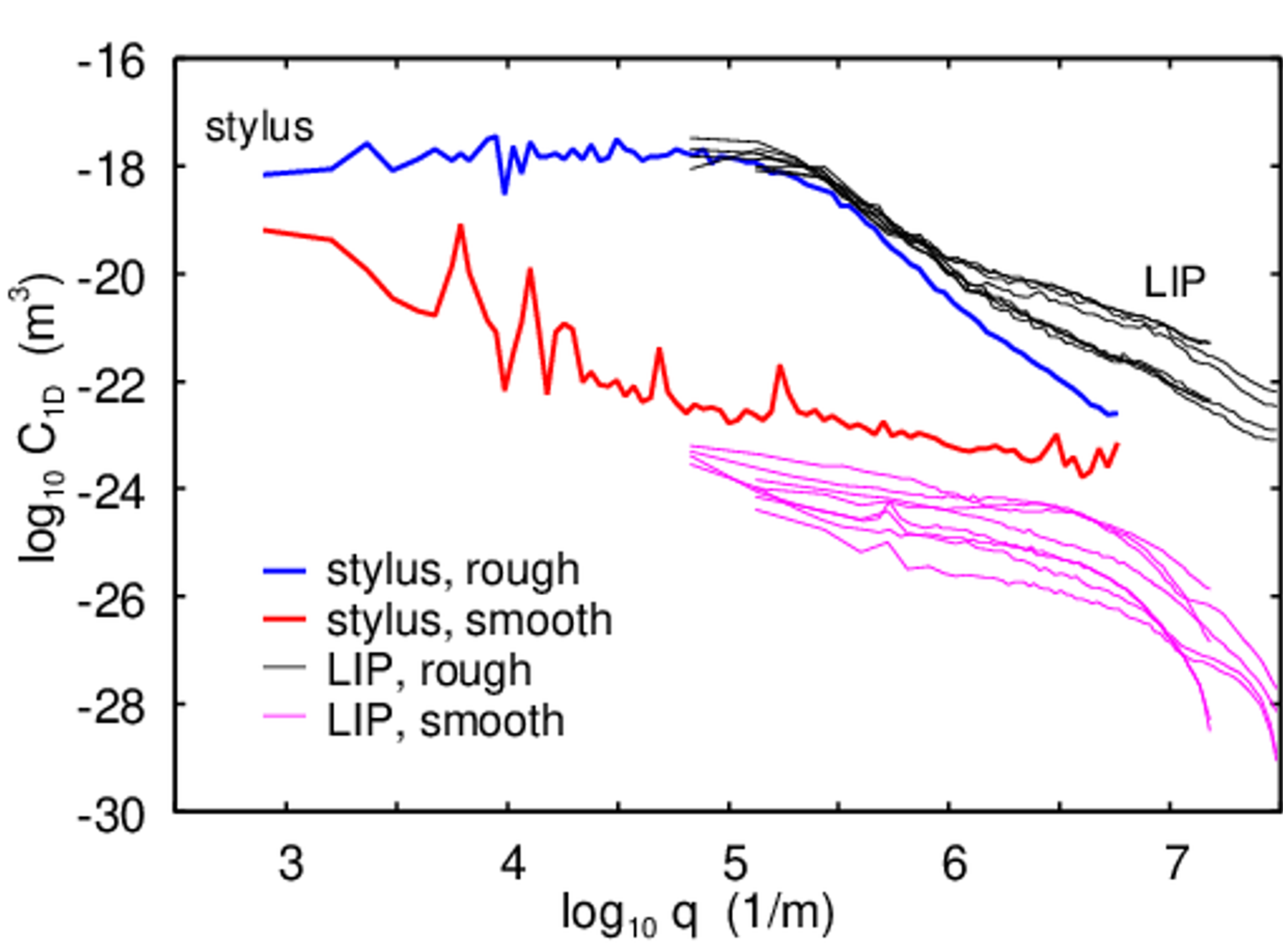}
\caption{\label{1logq.2logC.Julia.OP.1D.eps}
The 1D surface roughness power spectra of the smooth and rough surfaces calculated from topography data obtained using an
engineering stylus (red and blue curves), and Light Interferometry Profilometry (LIP). The LIP was performed in $480\times 640$ data points.
}
\end{figure}

A white and green 3D light interferometry profilometer Veeco NT-9100, 
with a $50\times$ magnification, was used to study the surface
topography of smooth and rough samples. 
The scanned area was of $94 \times 50 \ {\rm \mu m}^2$ with the pixel number $640 \times 480$.
Fig. \ref{1logq.2logC.Julia.OP.1D.eps} compares the $C_{\rm 1D}(q)$ results from the LIP to the output from
engineering stylus for the rough and smooth wafers. 

\vskip 0.1cm
{\bf White Light Profilomtry (WLP)}

\begin{figure}[!ht]
        \includegraphics[width=0.45\textwidth]{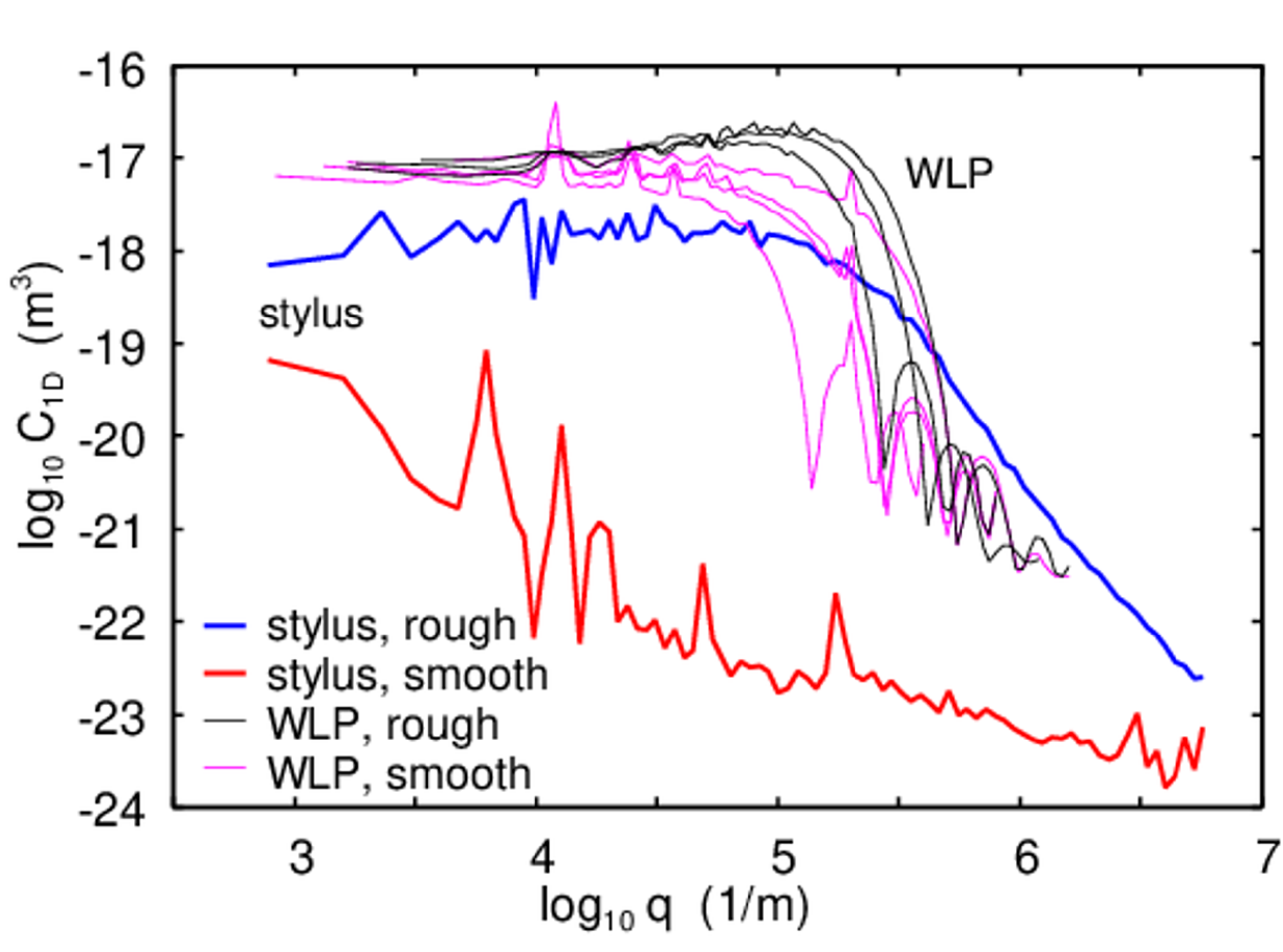}
\caption{\label{1logq.2logC.Julia.WL.1D.eps}
The 1D surface roughness power spectra of the smooth and rough surfaces calculated from topography data obtained using an
engineering stylus (red and blue curves), and White Light Profilometry (WLP). The WLP was performed in 
$1418 \times 1392$ data points.
}
\end{figure}

We have used a 3D Optical Profilometer (Keyence VR6200) to perform non-contact measurement 
with a stated resolution of $0.1 \ {\rm \mu m}$. Measurements were performed as follows:
(a) Single scan magnifications of 40, 80, and $120\times$ and scan
pixel number of $768 \times 1024$.
(b) Stitch mode at $80\times $  magnification at the scan pixel number 
$1392 \times 1418$ (rough surface) and $1108 \times 1283$ (smooth surface).
Fig. \ref{1logq.2logC.Julia.WL.1D.eps} compare the $C_{\rm 1D}(q)$ results from the WLP compared to the output
from engineering stylus for the rough and smooth wafers.

\vskip 0.3cm
{\bf 5 Another optical method for obtaining $C(q)$}

We have shown that standard optical methods in general
does not result in accurate surface roughness power spectra.
They may approximately describe the long wavelength roughness, and
may therefore result in good-looking topography pictures, but fail
for the short wavelength roughness. However, for very smooth surfaces
with roughness amplitude such that $k_0 h_{\rm rms} < 0.1$, where $k_0 = 2 \pi /\lambda_0$
is the wavenumber of the light used, an essential exact way 
exist for obtaining $C({\bf q})$ from light reflection data.

Based on the original work of Rayleigh \cite{Ray}, using first-order vector perturbation theory,
Rice \cite{Ric} and others have shown that for smooth enough surfaces the angular distribution of light
scattered from a rough surface depends on the surface roughness only via the power spectrum
$C({\bf q})$ where ${\bf q}$ is the change in the photon wavevector in the $xy$-plane between the
incident and reflected (scattered) photon beam:
$${\bf q} = {\bf k}_\parallel -{\bf k}^{\rm (i)}_\parallel = (k_x -k_x^{\rm (i)},k_y -k_y^{\rm (i)})$$
where ${\bf k}$ is the wavevector of the scattered beam.
The intensity of the light scattered into the solid-angle
$d\Omega = {\rm sin}\theta d\theta d\phi $ is given by \cite{Radar,More}
$${1\over I_{\rm i}} {dI \over d\Omega} d\Omega \approx 4k_0^4 \ {\rm cos} \theta_{\rm i} \ {\rm cos}^2 
\theta \ Q \ C_{\rm 2D}({\bf q}) d\Omega \eqno(16)$$
where $I_{\rm i}$ and $\theta_{\rm i}$ are the intensity of the incident beam and the angle of incidence, respectively, and
where the factor $Q$ depends on the
incident and scattering angles and on the dielectric constant 
of the solid. Note the Rayleigh blue-sky factor
$k_0^4$. 

Assuming small scattering angles, (16) simplifies to
$${1\over I_{\rm s}} {dI \over d\Omega} d\Omega \approx 4k_0^4 \ {\rm cos}^3 
\theta_{\rm i} \ C_{\rm 2D}({\bf q}) d\Omega \eqno(17)$$
where $I_{\rm s} = R I_{\rm i}$ is the intensity of the reflected beam ($R$ is the reflection factor and $I_{\rm i}$
the intensity of the incoming beam).

        \begin{figure}[!ht]
        \includegraphics[width=0.45\textwidth]{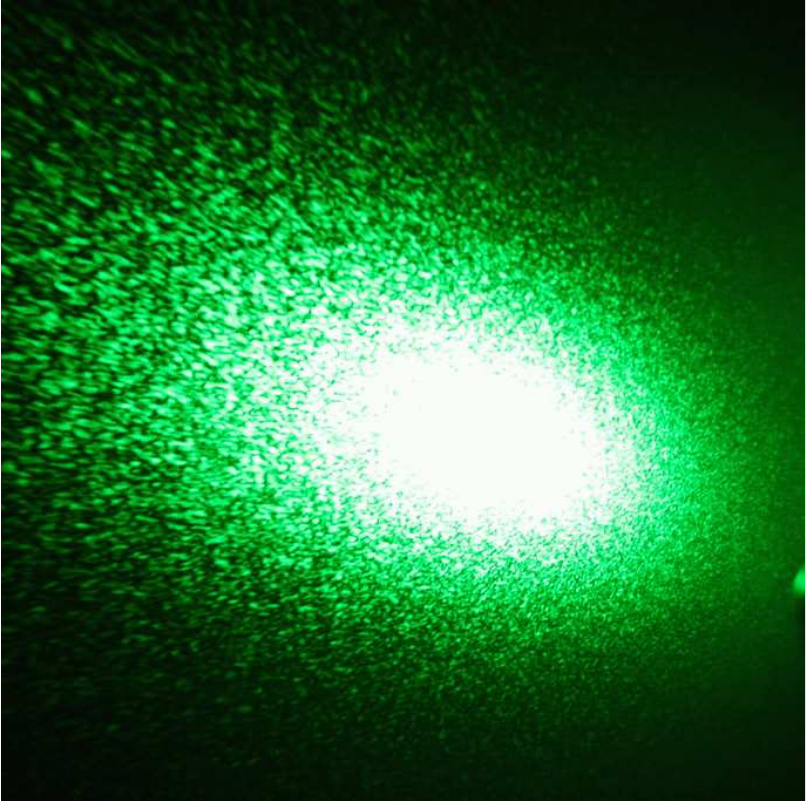}
\caption{\label{Speckle.ps}
Laser light is reflected from a rough surface. The random granular pattern (speckles) is due to light reflection
from the microscopic surface roughness and will look different when the laser beam is moved to another 
nominally identical surface region on the macroscopic body under study. 
The speckles disappear when the incident photon beam is incoherent.
}
\end{figure}

We have stated above that the $\langle .. \rangle$ in the definition of
$C({\bf q})$ stands for ensemble averaging. If no ensemble averaging is performed 
the $C({\bf q})$ obtained from a single surface area
will have some fine-scale noise reflecting the particular surface 
roughness occurring on the studied surface area. That is,
without ensemble averaging $C({\bf q})$ will have fine-scale speckle structure (noise) 
reflecting the surface area used. This speckle noise is well known 
to anybody who has observed laser light reflected from surfaces where
the term speckle refers to a random granular pattern that can be observed when a highly 
coherent light beam is diffusely reflected at a surface (see Fig. \ref{Speckle.ps}).
This phenomenon results from the interference of many different reflected portions of the incident 
beam with random relative optical phases. We note that if the incident beam is incoherent, speckle effects are
automatically averaged out.

One interesting limiting case of (16) is for a one-dimensional (grating-like) roughness profile, $z=h(x)$,
where
$$h({\bf q}) = {1\over (2\pi )^2} \int d^2x \ h({\bf x}) e^{-i{\bf q}\cdot {\bf x}} $$
$$= {1\over 2\pi } \int dx \ h(x) e^{-iq_x x} \delta (q_y) = h(q_x) \delta (q_y)$$ 
Using that
$$|\delta (q_y)|^2 = \delta (q_y) {1\over 2 \pi} \int dy \ e^{-iq_y y} = {L_0 \over 2 \pi} \delta(q_y)$$
where $A_0=L_0^2$ is the studied surface area, we get
$$C_{\rm 2D}({\bf q}) = {(2\pi )^2 \over A_0} |h({\bf q})|^2 $$
$$= {2\pi  \over L_0} |h(q_x)|^2 \delta(q_y) =C_{\rm 1D}(q_x)  \delta(q_y) $$
We substitute this in (16) and integrate over $\phi$. To evaluate the $\phi$-integral we use that 
${\rm sin} \phi_{\rm i}=0$ (we assume the wavevector of the incident beam is in the $xz$-plane) 
and $q_y=k_0 {\rm sin}\theta {\rm sin \phi}$ so that
$$\int_0^{2\pi} d\phi \ f(\phi) \delta (q_y) = \int_0^{2\pi} d\phi \ f(\phi) \delta (k_0 {\rm sin}\theta {\rm sin \phi}) $$
$$= \sum_{\phi =0, \pi}{f(\phi)\over |k_0 {\rm sin}\theta {\rm cos}\phi |} = {f(0)+f(\pi)\over k_0 {\rm sin}\theta}\eqno(18)$$
Here we have used that ${\rm sin}\phi =0$ for $\phi=0$ and $\phi=\pi$.
Using this result in (16) gives
$${1\over I_{\rm i}} {dI \over d\theta} d\theta \approx 8 k_0^3 \ {\rm cos} \theta_{\rm i} \ {\rm cos}^2 \theta \
Q \ C_{\rm 1D}(q_x) d\theta \eqno(19)$$
where $q_x =k_0 {\rm sin}\theta {\rm cos}\phi = \pm k_0 {\rm sin}\theta$ since $\phi=0$ or $\phi=\pi$. However,
$C_{\rm 1D}(q_x)=C_{\rm 1D}(-q_x)$ so we may use $q_x=k_0 {\rm sin}\theta$.
In (19) $Q$ is evaluated for $\phi=0$.
Note that the ${\rm sin}\theta$ factor in $d\Omega$ cancels against the same factor occurring in the denominator in (17).

If $h(x)= h_0 {\rm sin}(q_0 x)$ we get
$$C_{\rm 1D} (q_x) = {\pi \over 2} {h_0^2\over L_0} |\delta (q_x-q_0) - \delta (q_x+q_0)|^2 $$
$$= {h_0^2 \over 4} \left [ \delta (q_x-q_0) + \delta (q_x+q_0)\right ]$$
$$= {h_0^2 \over 4 k_0} \left [ \delta (q_x/k_0-q_0/k_0) + \delta (q_x/k_0+q_0/k_0)\right ]\eqno(20)$$
Using (19) and (20) gives the angular dependency of the scattered light from a sinus-grating.

Since the incident photon beam has the wavevector in the $xz$-plane we have $k^{\rm (i)}_x = k_0 {\rm sin}\theta_{\rm i}$
and since $k^{\rm (s)}_x =  k_0 {\rm sin}\theta_{\rm s}$ we get
$$q_x/k_0\pm q_0/k_0 = {\rm sin}\theta_{\rm s} - {\rm sin}\theta_{\rm i} \pm q_0/k_0$$
Thus, light incident on this surface profile will scatter in the form of a pair of first-order diffraction lines with positions
determined by the familiar grating equation
$${\rm sin}\theta_{\rm s} ={\rm sin}\theta_{\rm i}\pm q_0/k_0$$
where $q_0/k_0 = \lambda_0/d$,
where $d=2 \pi/q_0$ is the periodicity of the grating. The intensity of each of the grating lines are 
$\sim (k_0 h_0)^2$ relative to the incident intensity. This limiting case illustrates that the intensity
of the scattered light is determined by the square of the vertical amplitude of the roughness, while its angular spread
is determined by the spatial wavelength of the roughness. The latter is easy to understand as the momentum is transferred
to a photon by the grating is $\hbar {\bf q}$ to be compared to the incoming photon momentum
$\hbar {\bf k}_0$.

The method described above has been applied successfully to obtain information about surface roughness
using the scattering of laser light \cite{More} and X-rays \cite{XRR} from very smooth surfaces. However, 
we will now show that even the smooth surface studied above appears to be too rough for this method to be accurate.

        \begin{figure}[!ht]
        \includegraphics[width=0.45\textwidth]{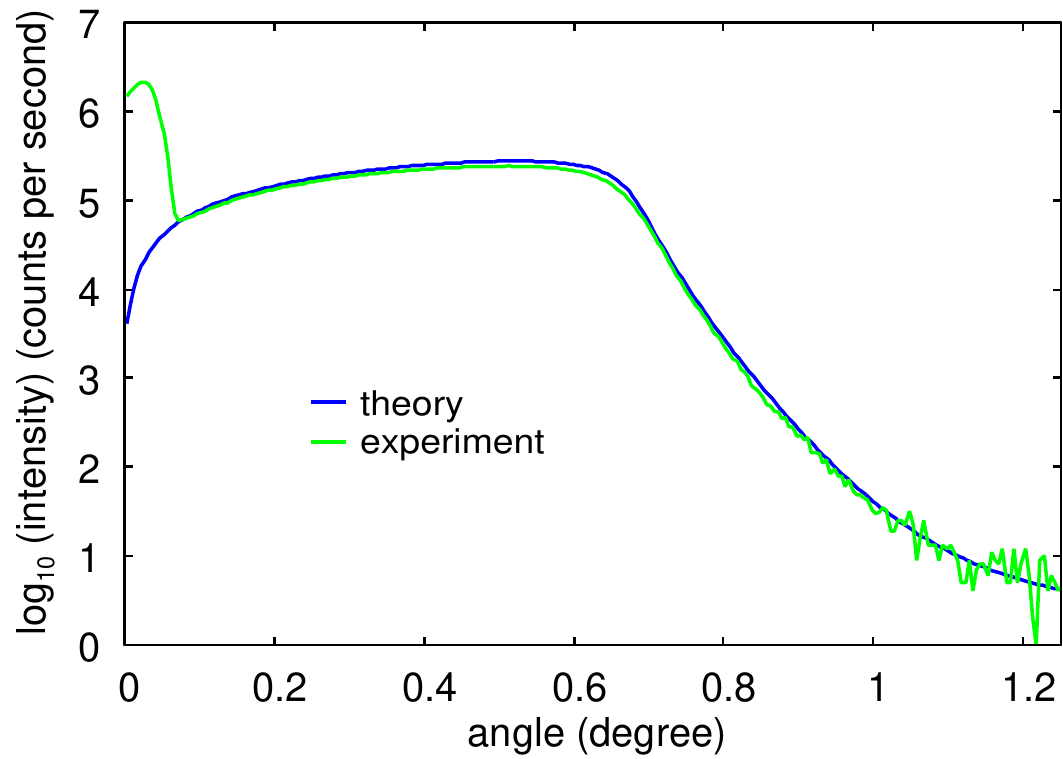}
\caption{\label{XRR.eps}
The measured data (green) and the fitted (blue) specular reflectivity. The results for the height fluctuations
and the assumed film thickness are given in the text.
}
\end{figure}

\vskip 0.1cm
{\bf X-Rays Reflectivity (XRR)}

One of the smooth samples, C13, was analyzed using a Panalytical Empyrean Multipurpose X-ray diffractometer. 
The measurement was taken using Cu ka of wavelength $1.546 \ {\rm \AA}$. The incident beam was shaped using a 
Bragg-Brentano mirror and the reflected beam were detected using a Pixel-3D  area detector in 
receiving mode with 1 of 255 channels active. Symmetric gonio scans were taken 
from 0 to $4^\circ$ of $2\theta$ with a step size of $0.005^\circ$ $2\theta$ and measurement speed of 
$0.44 \ {\rm s/step}$. The data was fitted in 
X'Pert Reflectivity (V1.3) software to estimate the mass density and the rms-roughness.
Figure \ref{XRR.eps} shows the measured data (green) and the fitted model (blue). 
Assuming a $1.5 \ {\rm \mu m}$ thick CrN film, the density obtained from the fitting was 
$5.85 \ {\rm gm/cm^3}$ and the roughness was $4.4 \ {\rm nm}$. 
Since the X-ray beam width was $6 \ {\rm mm}$, covering almost the full wafer dimension along the beam direction, 
the rms-roughness should be compared with the one obtained from the engineering stylus which is $\sim 20 \ {\rm nm}$. 
AFM and XRR roughness measurements found in the literature show good
agreement for surfaces with atomic-scale roughness and up to a few nanometers \cite{XRR}. 
In our case, the smooth wafers appear too rough for the XRR method to be accurate.

        \begin{figure}[!ht]
        \includegraphics[width=0.45\textwidth]{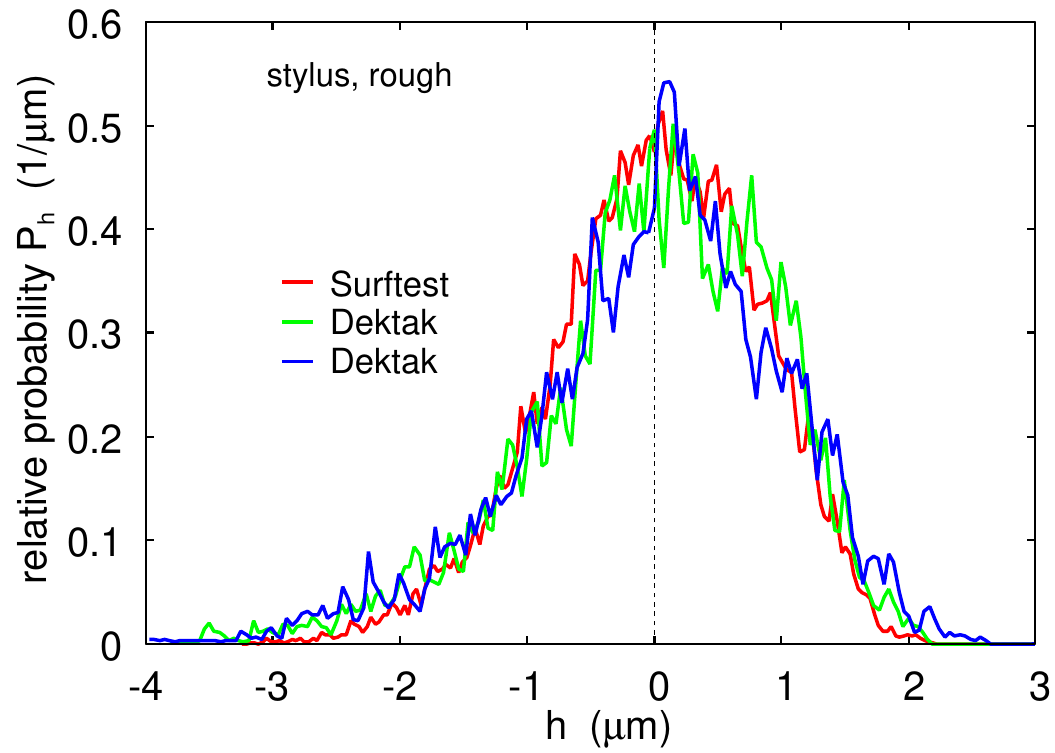}
\caption{\label{1h.2Ph.redJuelich.BlueGreenBD.eps}
The height probability distribution for the rough surface
obtained using the engineering stylus instruments.
}
\end{figure}

        \begin{figure}[!ht]
        \includegraphics[width=0.45\textwidth]{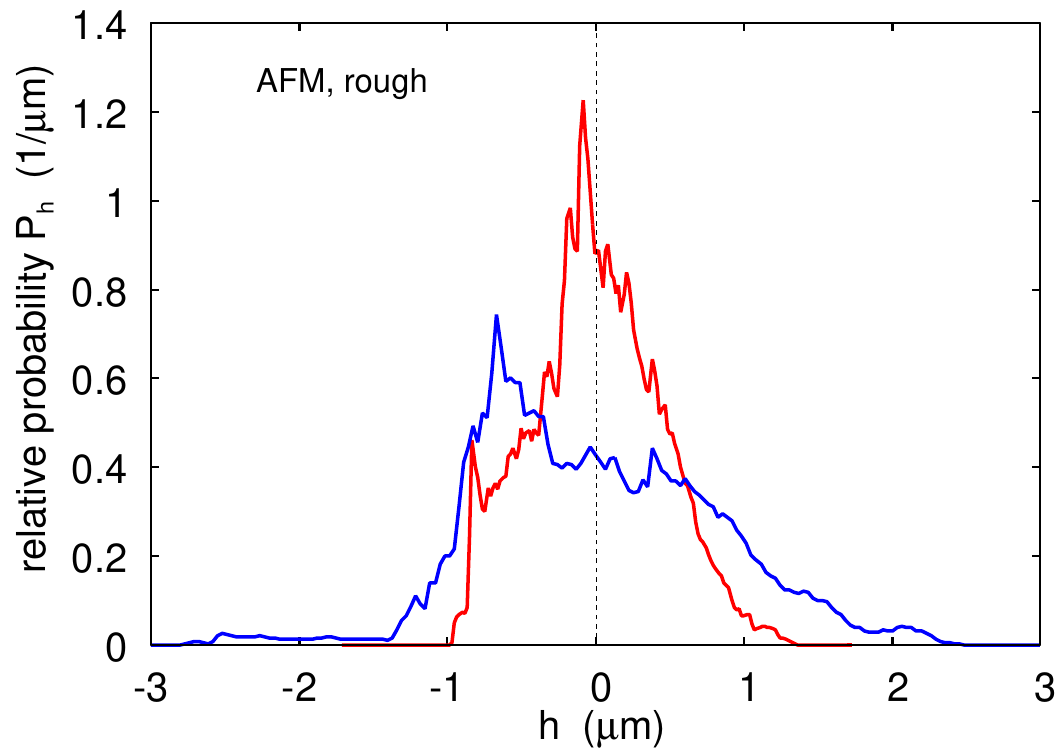}
\caption{\label{1h.2Ph.AFM.eps}
The height probability distribution for the rough surface
obtained using AFM.
}
\end{figure}

        \begin{figure}[!ht]
        \includegraphics[width=0.45\textwidth]{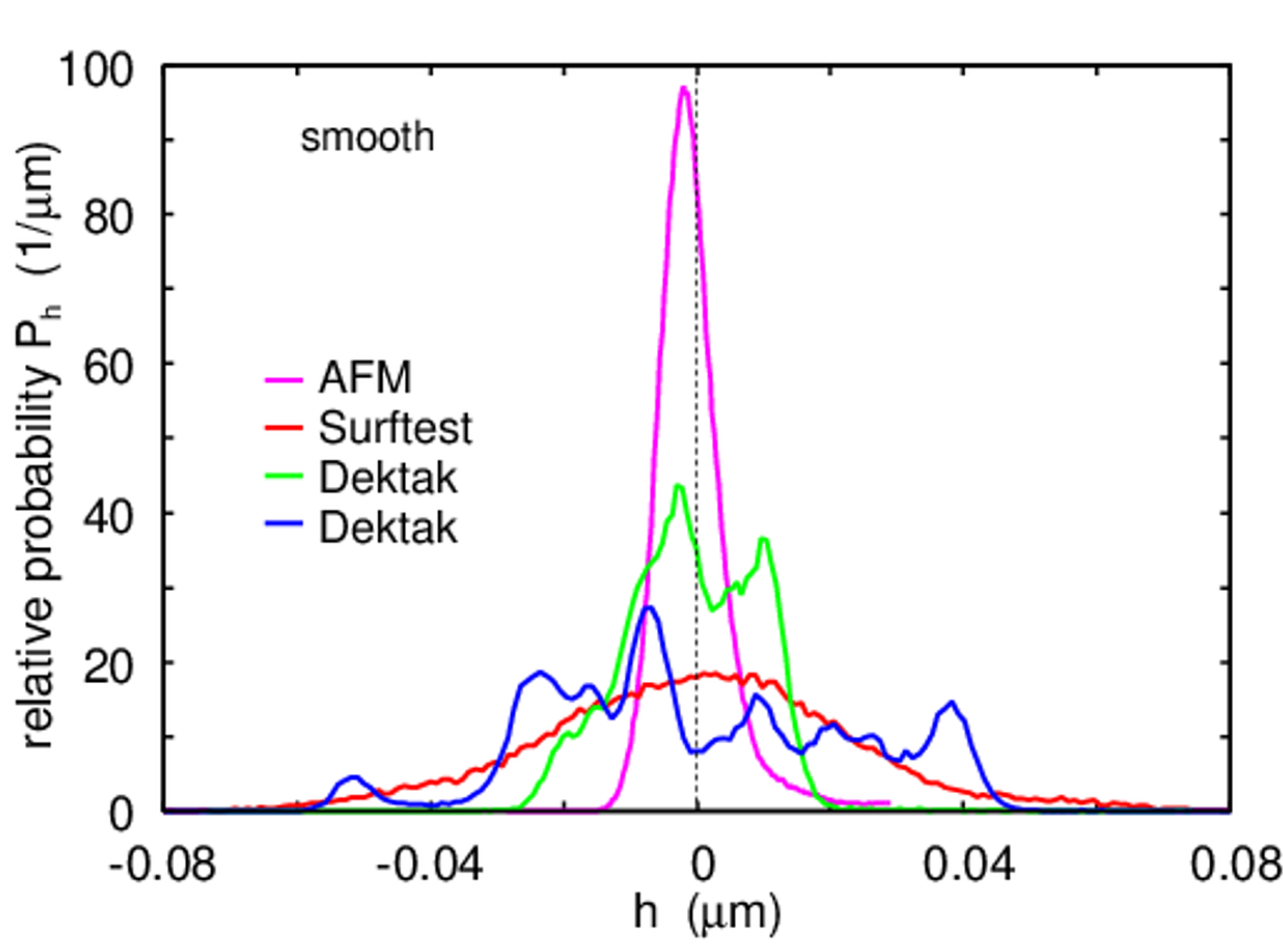}
\caption{\label{1h.2Ph.smooth.redJulia.pinkAFM.otherBD.eps}
The height probability distribution for the smooth surface
obtained using AFM and the engineering stylus instruments.
}
\end{figure}

\vskip 0.3cm
{\bf 6 Surface roughness parameters}

The most complete information about surface roughness
is the height probability distribution $P_h$ and the surface
roughness power spectrum $C({\bf q})$. In Sec. 4 we showed the power spectra, and in Fig. \ref{1h.2Ph.redJuelich.BlueGreenBD.eps}, \ref{1h.2Ph.AFM.eps} 
and \ref{1h.2Ph.smooth.redJulia.pinkAFM.otherBD.eps} we show the height probability distributions. 
Note that for the rough surface, all the stylus measurements give very similar
$P_h$ as expected when a large roll-off region is included in the measurements \cite{my,my1}.
This is not the case for the AFM data because it does not include a roll-off region. 
For the smooth surface, the Dektak stylus instrument gives very strongly fluctuating $P_h$ while 
the Surftest gives a smoother curve which may, in part, result from the fact
that the Dektak values are averages over only 3 lines.

In most engineering applications just one or two parameters 
(numbers) are used, typically the rms roughness $h_{\rm rms}$ (or the arithmetic average) or maximum height of roughness ${\rm Rz}$, 
are given to characterize the surface roughness. 
A very large number of surface roughness parameters have been defined \cite{many}
but in our opinion, only a few are really useful, namely 
$h_{\rm rms}$ and rms slope $\xi$, which can be
obtained as integrals involving $C({\bf q})$, and the
the skewness $S_{\rm sk}$ and the kurtosis $S_{\rm ku}$, which can be
obtained as integrals involving $P_h$ (see Sec. 2).
The latter two parameters are important because they indicate to what extent
a surface is randomly rough since in that case $S_{\rm sk}=0$
and $S_{\rm ku}=3$. (Note: if there is no roll-off region it is important to perform an
average over many independent measurements (ensemble average) as otherwise the $P_h$,
and hence also $S_{\rm sk}$ and $S_{\rm ku}$, will 
fluctuate widely between different measurements.) 
The rms roughness $h_{\rm rms}$ is important for the
average surface separation when two elastic solids are squeezed into contact while
the rms slope $\xi$ is important in a large number of applications, e.g., for 
determining the area of real contact between two elastic solids. In most cases
$h_{\rm rms}$ depends mainly on the long wavelength roughness and is hence easy to measure using,
e.g., an engineering stylus instrument if measured over a line long enough to cover the roll-off region
of the power spectrum. The rms slope $\xi$ depends on the short wavelength roughness, and to
determine it, one may need to combine AFM with engineering line scan measurements to cover
all relevant length scales.

\begin{table}[hbt]
\caption{The rms roughness $h_{\rm rms}$ (in ${\rm \mu m}$) and rms slope $\xi$ as obtained from
engineering stylus and AFM data}
\label{Hh}
\begin{center}
      \begin{tabular}{@{}|l||c|c|c|c|c|@{}}
\hline
method & $h_{\rm rms}$  & $\xi$\\
\hline
\hline
AFM (rough) & $0.759$, $0.422$ & $0.349$, $0.327$\\
\hline
Surftest (rough) & $0.819$ & $0.258$\\
\hline
Dektak (rough) & $1.003$, $0.961$ & $0.346$, $0.334$\\
\hline
\hline
AFM (smooth) & $0.00622$ & $0.186$\\
\hline
Surftest (smooth) & $0.0235$ & $0.0333$\\
\hline
Dektak (smooth) & $0.0245$, $0.0186$ & $0.0465$, $0.0737$\\
\hline
\end{tabular}
   \end{center}
\end{table}

\begin{table}[hbt]
\caption{The skewness $S_{\rm sk}$ and the kurtosis $S_{\rm ku}$ as obtained from
engineering stylus and AFM data}
\label{Ss}
\begin{center}
      \begin{tabular}{@{}|l||c|c|c|c|c|@{}}
\hline
method & $S_{\rm sk}$  & $S_{\rm ku}$ \\
\hline
\hline
AFM (rough) & $0.173$, $0.124$ & $3.194$, $2.644$\\
\hline
Surftest (rough) & $-0.378$ & $3.122$\\
\hline
Dektak (rough) & $-0.580$, $-0.719$ & $3.682$, $3.768$\\
\hline
\hline
AFM (smooth) & $2.683$ & $13.708$\\
\hline
Surftest (smooth) & $-0.399$ & $15.584$\\
\hline
Dektak (smooth) & $0.0720$, $9.817$ & $10.762$, $194.456$\\
\hline
\end{tabular}
   \end{center}
\end{table}

In table \ref{Hh} we show the rms roughness $h_{\rm rms}$ (in ${\rm \mu m}$) 
and rms slope $\xi$, and in table \ref{Ss} we show the skewness $S_{\rm sk}$ and the kurtosis $S_{\rm ku}$ calculated from
engineering stylus and AFM data (the same data as used in Fig. \ref{1logq.2logC.Julia.and.BD.stulus.repeat.eps}
and Fig. \ref{1logq.2logC1D.AFMandJulia.eps}). These quantities have been obtained using the measured topography data over different length scales and this must be taken into account when comparing the results using the different methods. For the rough surface, the values from different methods are similar but for the smooth surface, they vary strongly even when comparing two measurements using the same method. These strong fluctuations in the parameter values
may be related to the strong fluctuations that can be observed in the power spectra for the smooth surface (see Fig. \ref{1logq.2logC.Julia.and.BD.stulus.repeat.eps}). The engineering  stylus data of the rough surface shows negative skewness while the 
AFM gives positive skewness. This indicates that the roughness in the roll-off region, which is not probed in the AFM measurements, 
may have different sign from the skewness in the self-affine fractal region.

        \begin{figure}[!ht]
        \includegraphics[width=0.45\textwidth]{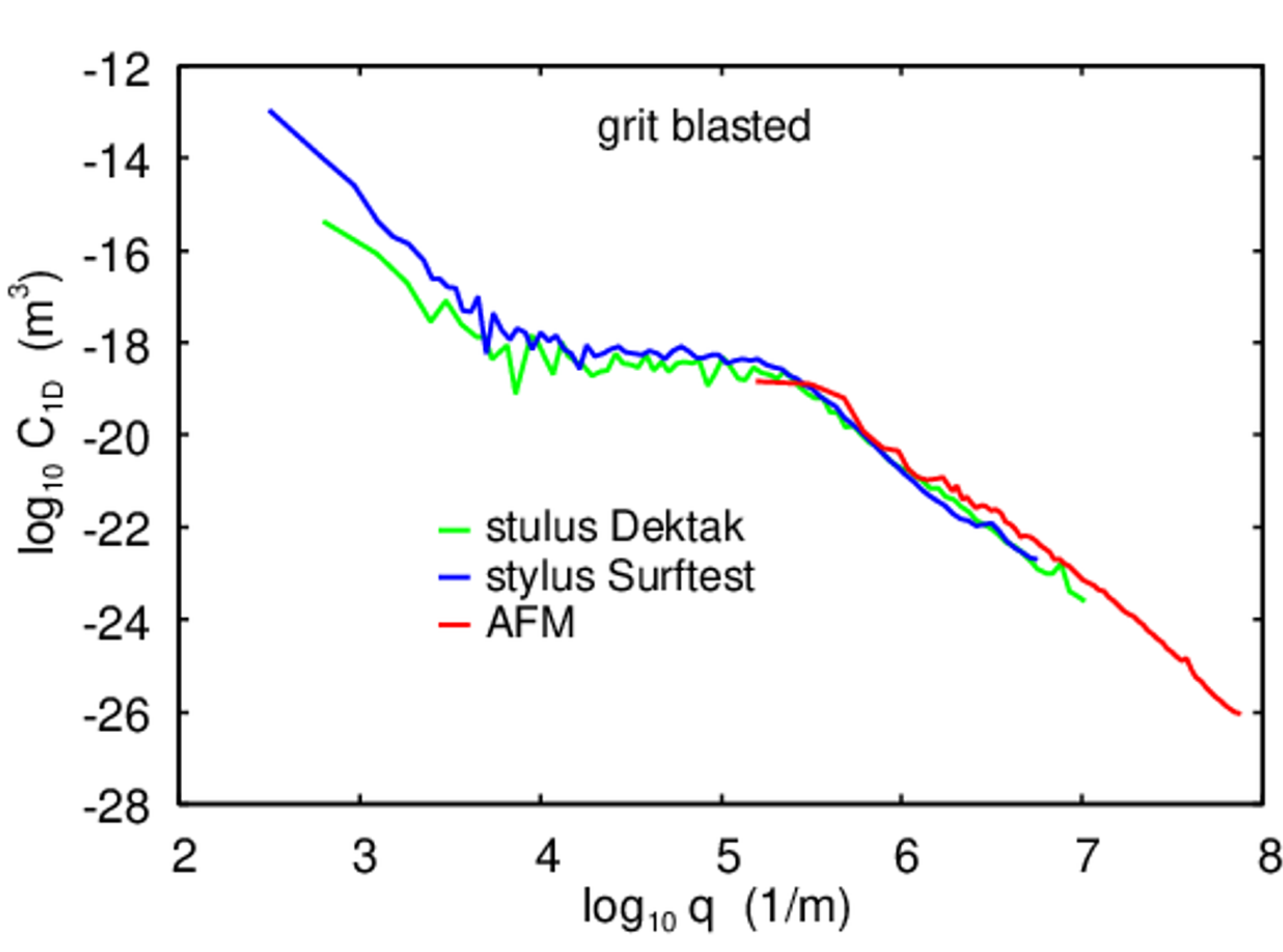}
\caption{\label{1logq.2logC.GRIT.eps}
The power spectrum of a grit blasted nickel surface (arithmetic average ${\rm Ra} = 32 \ {\rm \mu inch}$)
supplied by Gar Electroforming \cite{Gar} as a calibration standard.
}
\end{figure}

        \begin{figure}[!ht]
        \includegraphics[width=0.45\textwidth]{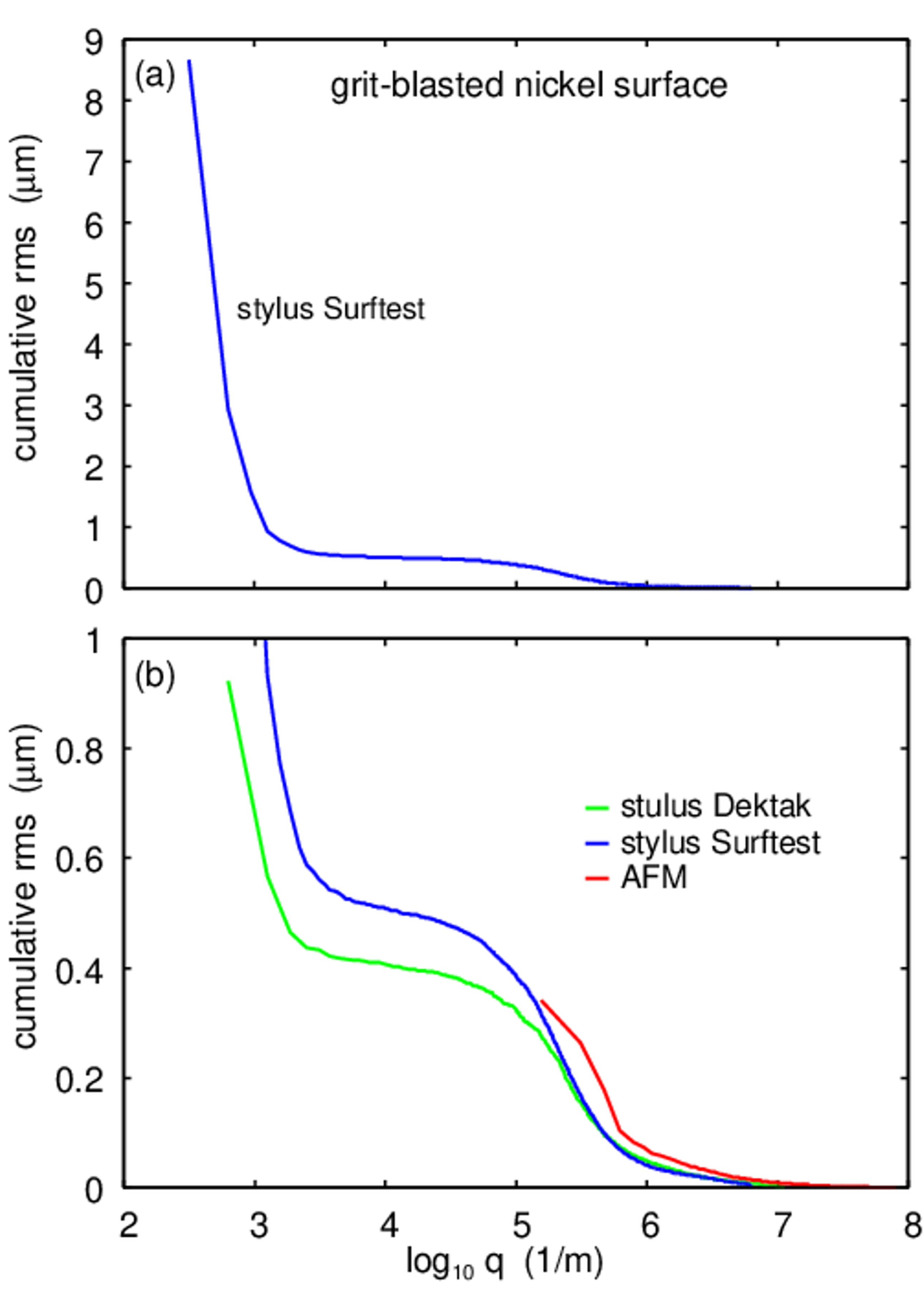}
\caption{\label{1logq.2rms.cumulative.GRIT.eps}
The cumulative rms-roughness $h_{\rm rms}$ as a function of the small cut-off wavenumber. In the calculation of
$h_{\rm rms}(q)$ only the roughness components with wavenumber between $q$ and the large cut-off wavenumber  $q_1$
(which equals $8\times 10^6 \ {\rm m}^{-1}$ for the stylus measurement) is included in the calculation. 
(a) results using the Surftest stylus, and (b) the same as in (a) but including results from the Dektak stylus and the AFM. 
Results are obtained using (19) with
the power spectra shown in Fig. \ref{1logq.2logC.GRIT.eps}. 
}
\end{figure}

        \begin{figure}[!ht]
        \includegraphics[width=0.45\textwidth]{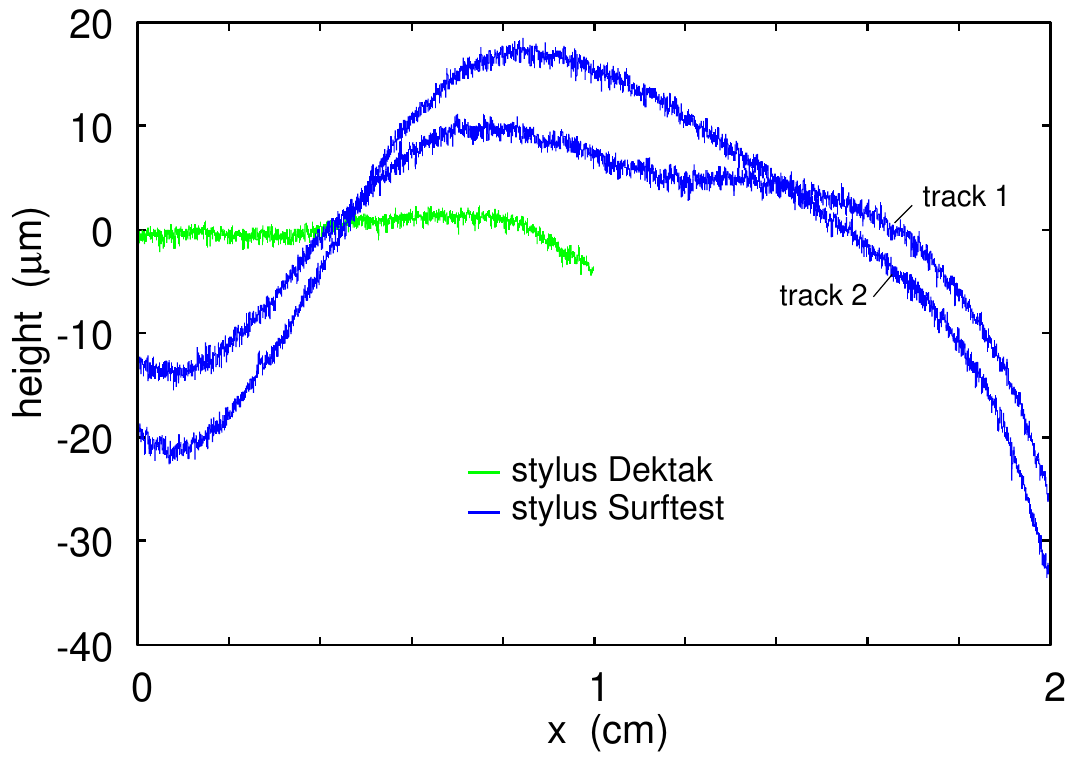}
\caption{\label{1x.2height.GRIT.eps}
Line scan tracks obtained using the Surftest stylus (blue) and the Dektak stylus (green) 
after removing the average slope of the measured data but not the macroscopic curvature as the grit-blasted surface was nominally flat.
The line scan data was used to calculate the
power spectra in Fig. \ref{1logq.2logC.GRIT.eps}. 
}
\end{figure}

\vskip 0.3cm
{\bf 7 On the use of calibration and filters}

Calibration is the comparison between measurement values delivered by a device under test with those of a calibration standard of known accuracy. For calibration of topography instruments, many calibration standards exist but most of them just specify the rms roughness (or rather the arithmetic average) and sometimes the lateral spacing distance in a periodic profile e.g., a sinus-like profile. 
To illustrate the problem with this, we show in Fig. \ref{1logq.2logC.GRIT.eps} 
the power spectrum of a grit-blasted nickel surface which is used as a calibration standard. The supplier
of this calibration standard states that it has the arithmetic average roughness ${\rm Ra} = 32 \ {\rm \mu inch}$
or about $0.81 \ {\rm \mu m}$. Assuming that the rms-roughness is a factor of $(\pi/2)^{1/2}$ larger than ${\rm Ra}$
as expected for random roughness, this gives $h_{\rm rms} \approx 1 \ {\rm \mu m}$. However, from the $20 \ {\rm mm}$
long stylus topography measurement, we obtained $h_{\rm rms} \approx 10 \ {\rm \mu m}$. This difference from what is expected
must be due to the absence of filtering in our study. Fig. \ref{1logq.2logC.GRIT.eps} shows that the roughness
on the calibration probe consists of short wavelength roughness separated by a roll-off region from the long wavelength
roughness (usually denoted waviness), which extends to the linear size of the sample. Since the rms roughness $h_{\rm rms}$ 
usually is determined by the most long wavelength roughness components, removing the waviness may result in a much smaller
rms-roughness in better agreement with the quoted roughness value. 

To prove this assumption, in Fig. \ref{1logq.2rms.cumulative.GRIT.eps} we show
the cumulative rms roughness $h_{\rm rms}(q)$ as a function of the small cut-off wavenumber $q$:
$$h^2_{\rm rms}(q) = 2 \pi \int_q^{q_1} dq \ q C_{\rm 2D}(q)= 2 \int_q^{q_1} dq \ C_{\rm 1D}(q)\eqno(21)$$
Here $q_1$ is the largest wavenumber for which $C(q)$ is calculated. Fig. \ref{1logq.2rms.cumulative.GRIT.eps}(a) is obtained
from the Surftest stylus, and shows that as $q$ decreases towards $q_0$ (the smallest wavenumber for which $C(q)$ was measured) $h_{\rm rms}(q)$ increases to
$\approx 9  \ {\rm \mu m}$. The most useful definition of the roughness amplitude excluding the waviness roughness region
would be the value of $h_{\rm rms}(q)$ in the roll-off region which is about $\sim 0.5 \ {\rm \mu m}$ 
[see Fig. \ref{1logq.2rms.cumulative.GRIT.eps}(b)] which is somewhat smaller than the quoted
roughness. This indicates that the quoted roughness must have been obtained with a filter removing most of the waviness and roughness.

Figure \ref{1logq.2rms.cumulative.GRIT.eps}(b) also shows the cumulative rms roughness obtained from the power spectra
of the AFM and Dektak stylus instruments. Note that the Dektak stylus gives a smaller rms-roughness for small wavenumber $q$
than the Surftest stylus. We attribute this to the shorter line scan length in the Dektak measurement. Thus, in calculating
the power spectra and cumulative roughness we have only removed the slope (tilt) of the measured data, but not the macroscopic curvature as the grit blasted surface was nominally flat. This results in a larger long-range variation in the height for the Surftest stylus data when compared to the Dektak stylus data. This can be supported by Fig. \ref{1x.2height.GRIT.eps}, where the line scan tracks obtained using the Surftest stylus (blue) and the Dektak stylus (green) used to calculate the power spectra in Fig. \ref{1logq.2logC.GRIT.eps} are shown. 

Removing the waviness using a filter is a very arbitrary and not useful approach as in some applications the
long-wavelength roughness may be very important. This is often the case for the leakage of metallic seals, adhesion, or electric and thermal contact resistance. If information about the surface rms-roughness amplitude is important one should instead first calculate the power spectrum $C(q)$ (without filtering!!), from which one can obtain the rms-roughness [using (21)] including all relevant length scales. Similar length-scale dependent
quantities, such as the rms slope, can also be obtained from integrals over $C(q)$ similar to (21).

        \begin{figure}[!ht]
        \includegraphics[width=0.45\textwidth]{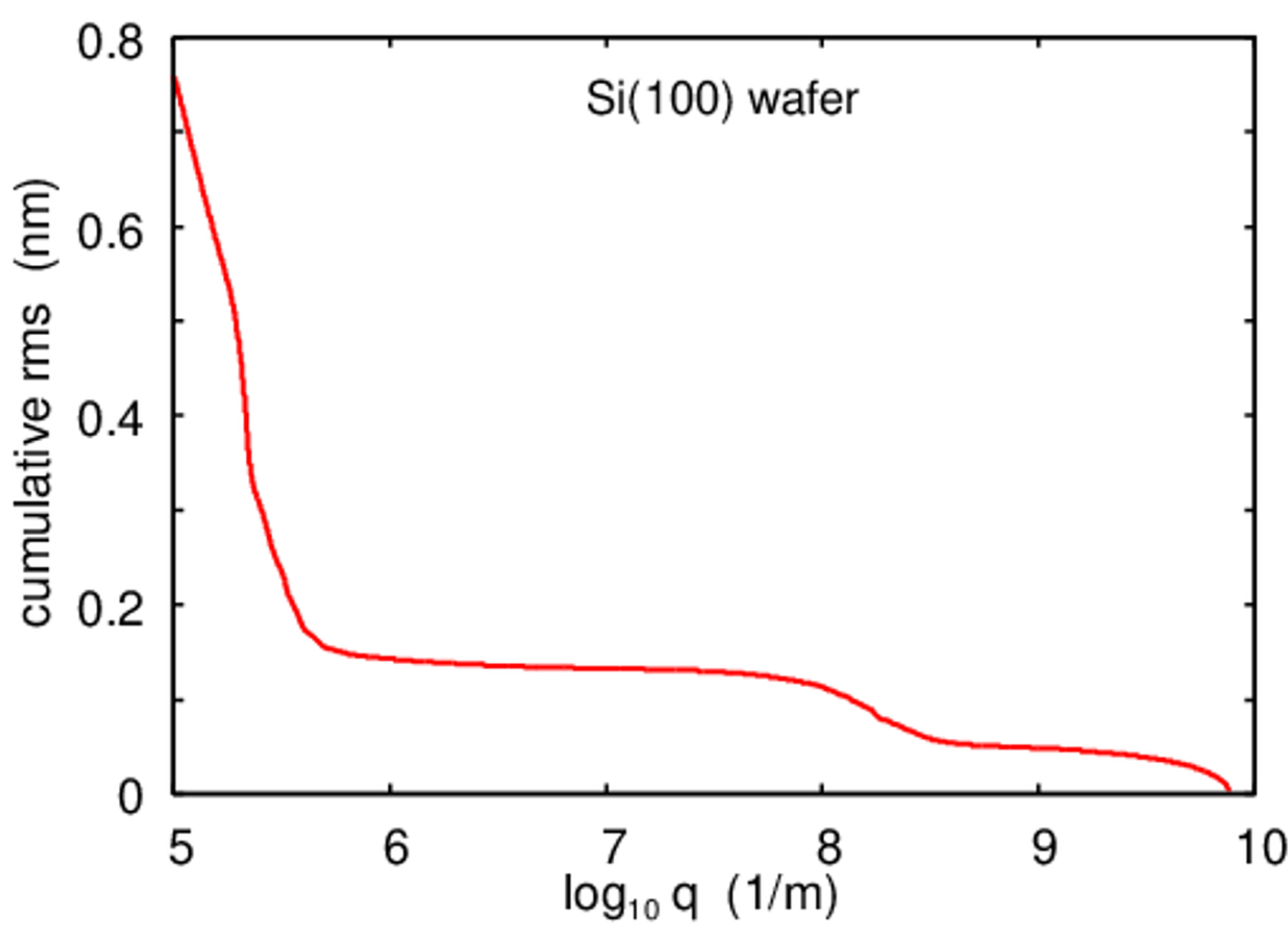}
\caption{\label{1logq.2logC.Si100.move.eps}
The cumulative rms-roughness $h_{\rm rms}$ as a function of the small wavenumber 
cut-off based on AFM measurements from a Si(100) wafer \cite{Mate}. 
In the calculation of
$h_{\rm rms}(q)$ only the roughness components with wavenumber between $q$ and the large wavenumber cut-off $q_1$
(which equals $8\times 10^9 \ {\rm m}^{-1}$) is included in the calculation. 
}
\end{figure}

Fig. \ref{1logq.2logC.GRIT.eps} shows good agreement between the AFM and the two stylus measurements, where the small difference (for a small wavenumber) between the two stylus results can be attributed to the different scan lengths. We have observed a similar agreement between the AFM and the stylus measurements for three other calibration standards where the surfaces were produced by lapping, grinding, and turning. In these cases, the stated $\rm Ra$ roughness is $0.05$, $0.05$, and $0.4 \ {\rm \mu m}$, respectively. But the roughness amplitudes we obtained including the waviness were much higher. 
A waviness region will occur on all smooth surfaces because it is nearly impossible to produce surfaces that are perfectly flat at the macroscopic length scale. This is the case even for wafers used in wafer bonding. Thus in Fig. \ref{1logq.2logC.Si100.move.eps} we show the cumulative rms roughness with decreasing small cut-off wavenumber for a Si(100) wafer obtained from AFM measurements. In Ref. \cite{Mate} it was shown that the wafer waviness had negligible influence on wafer bonding, but in other applications, waviness can be very important.

        \begin{figure}[!ht]
        \includegraphics[width=0.45\textwidth]{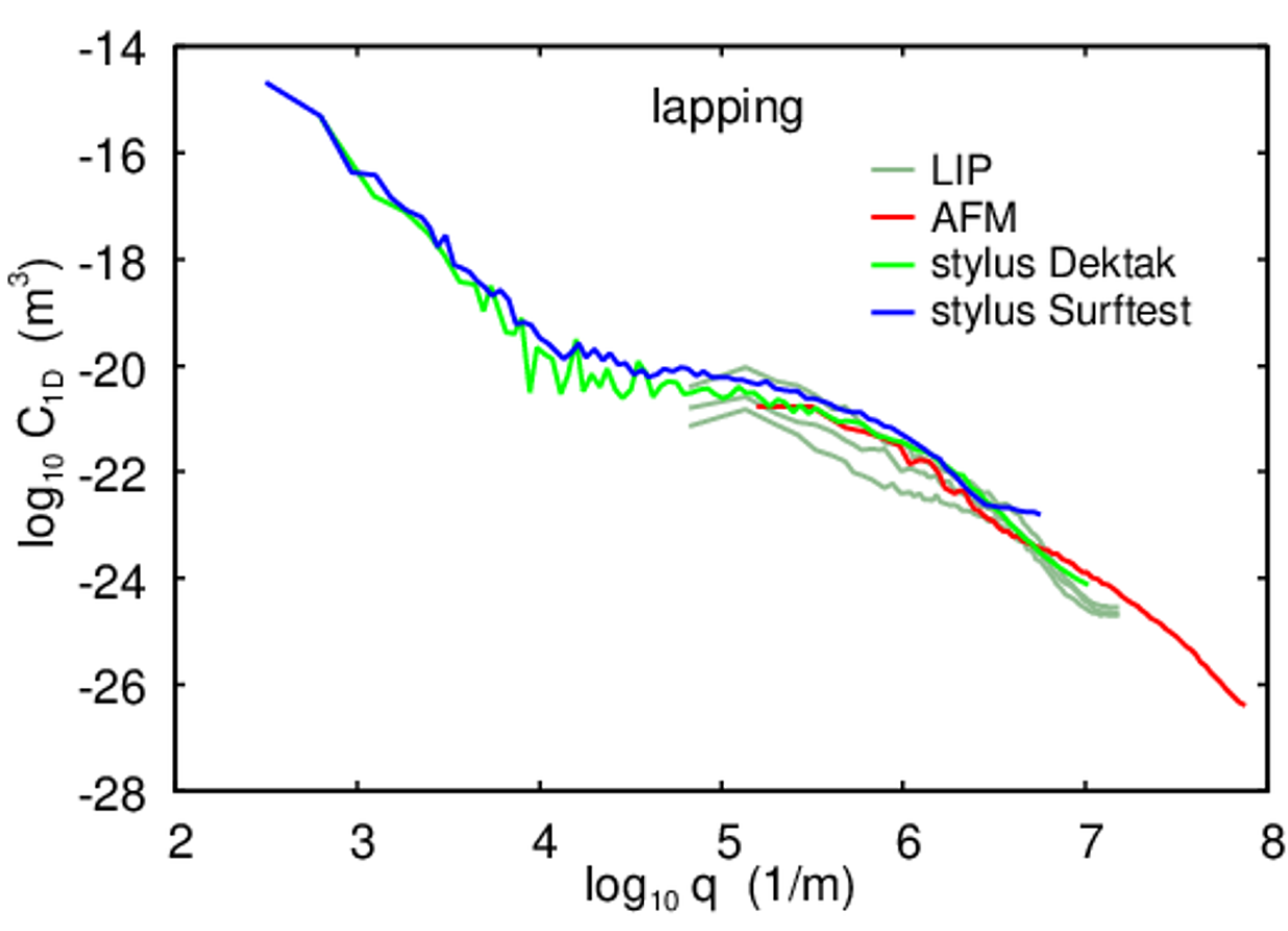}
\caption{\label{1logq.2logC1D.lapping.all.eps}
The power spectrum of a lapped nickel surface (arithmetic average ${\rm Ra} = 0.05 \ {\rm \mu m}$)
(FLEXBAR Model No. 16008, Surface Roughness Standards Set 800-879-7575). 
}
\end{figure}

        \begin{figure}[!ht]
        \includegraphics[width=0.3\textwidth]{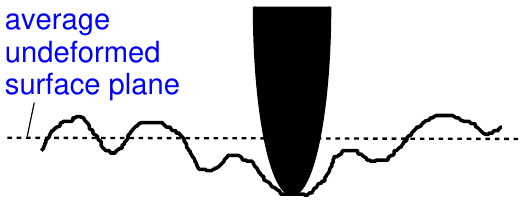}
\caption{\label{TipDeformPicBD.eps}
The stylus tip deforms the material elastically but if the deformations
are the same everywhere as the tip is scanned over the surface,
the measured topography will be the same as if the substrate is rigid.
}
\end{figure}

        \begin{figure}[!ht]
        \includegraphics[width=0.45\textwidth]{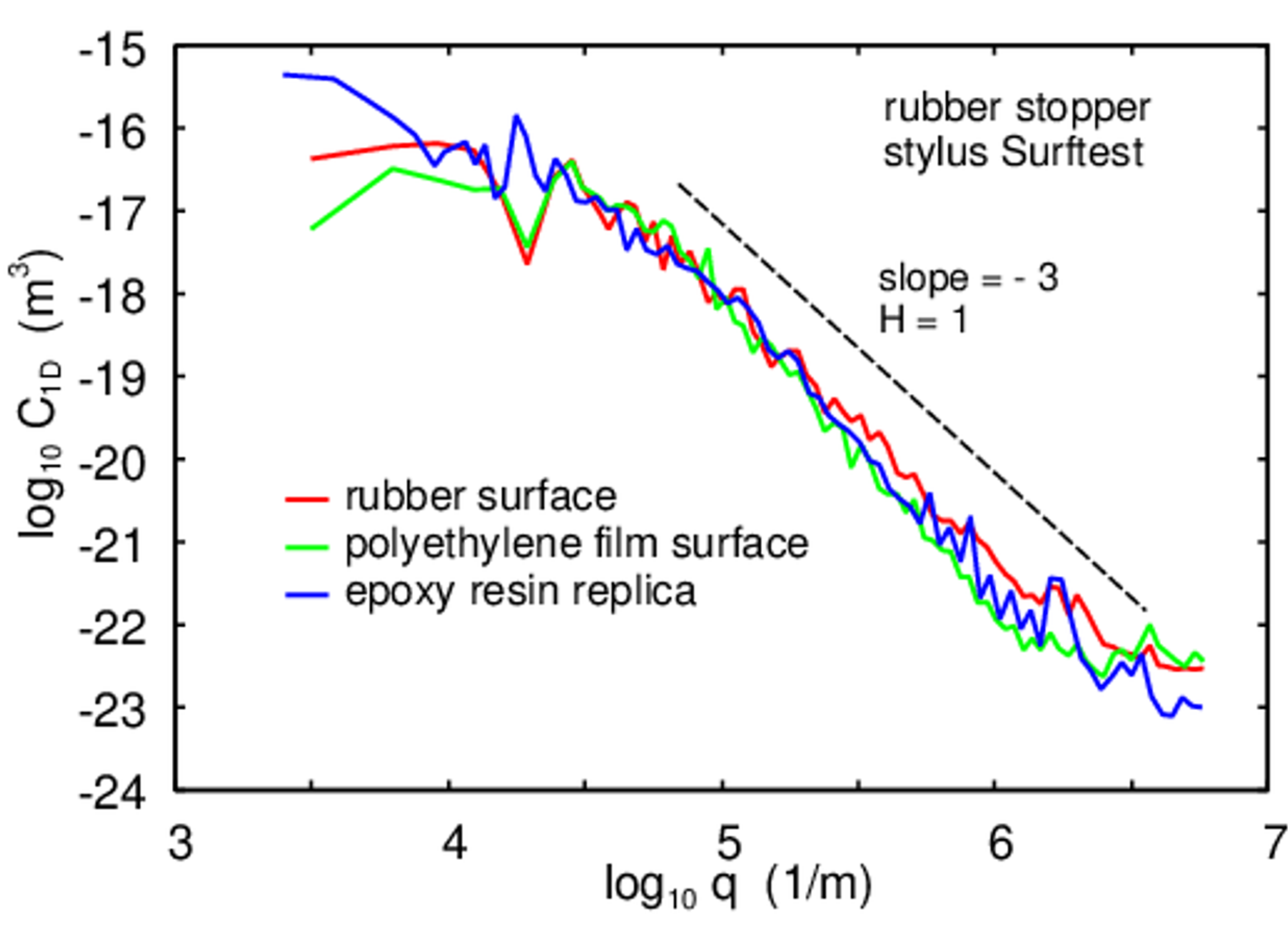}
\caption{\label{1q.2logC.BDStopperRedRubberGreenPlasticBlueReplica.eps}
The surface roughness topography of a rubber stopper used in a syringe. One part of the stopper
is covered by a $\sim 10 \ {\rm \mu m}$ thick polyethylene film, but the roughness on the rubber part (red curve)
is nearly the same as on the polyethylene film (green curve). This shows that the roughness on both surfaces results from the
steel mold. A replica of the stopper surface made from an elastically stiff epoxy resin 
has the same surface roughness power spectrum (blue curve)
as the other two surfaces.  This shows that tip-induced elastic deformation of the rubber surface has a negligible
influence on the measured topography (see Fig. \ref{TipDeformPicBD.eps}).
}
\end{figure}

\vskip 0.3cm
{\bf 8 Stylus measurements on soft materials}

For stylus measurements on soft solids, e.g., soft rubber materials, the tip-substrate force can effectively modify the real surface topography, as observed in Ref. \cite{Julia} for silicone rubber (PDMS) where stick-slip occurred. However, replicas made with elastically stiff (glassy) polymer on soft (or hard) originals can be used to study the topography via stylus instruments.

Fig. \ref{TipDeformPicBD.eps}
shows schematically how a stylus tip may deform the surface of an elastically soft material. This could result in a measured topography that differs from the real one. However, if the deformations are the same everywhere as the tip is scanned over the surface, the measured topography will be the same as if the substrate is rigid. To prove it, we have measured the surface topography of a soft (unfilled) rubber compound.

Fig. \ref{1q.2logC.BDStopperRedRubberGreenPlasticBlueReplica.eps} shows the surface roughness topography of a rubber stopper used in a syringe. The stopper was produced by injection molding with one part of the surface covered by a thin ($\sim 10 \ {\rm \mu m}$ thick) polyethylene film. The roughness on the rubber part (red curve) is nearly the same as on the polyethylene film (green curve) which indicates that the roughness on both surfaces results from the
steel mold. A replica of the stopper surface made from an elastically stiff epoxy resin has the same power spectrum (replica of the rubber part, blue curve) as the other two surfaces.  This shows that tip-induced elastic deformations of the rubber surface have a negligible influence on the measured topography.

A similar method was employed in the study referenced in \cite{Julia}, where the topography of smooth glass and silicone rubber replicas were measured, revealing significant differences. The discrepancy was attributed to stick-slip phenomena, likely due to adhesion between the measurement tip and the silicone rubber. Compared to the current study, the rubber used for the stopper has an elastic modulus $\sim 1.5$ times that of the silicone rubber, which is not significantly stiffer. Consequently, the difference in elastic modulus is unlikely to be the primary cause of the observed discrepancy. However, the silicone rubber replica in \cite{Julia} was cast from a very smooth surface, unlike the rougher surface of the rubber stopper. It is known that sufficient surface roughness can eliminate macroscopic adhesion \cite{kill}, where relevant roughness for this adhesion elimination corresponds to the roughness with a wavelength shorter than the width of the rubber-tip contact region, approximately $1 \ \mu {\rm m}$. We hypothesize that this is the main distinction between the two systems. To further substantiate this hypothesis, additional measurements were conducted on PDMS replicas of both smooth and sandblasted PMMA and silica glass surfaces.

        \begin{figure}[!ht]
        \includegraphics[width=0.45\textwidth]{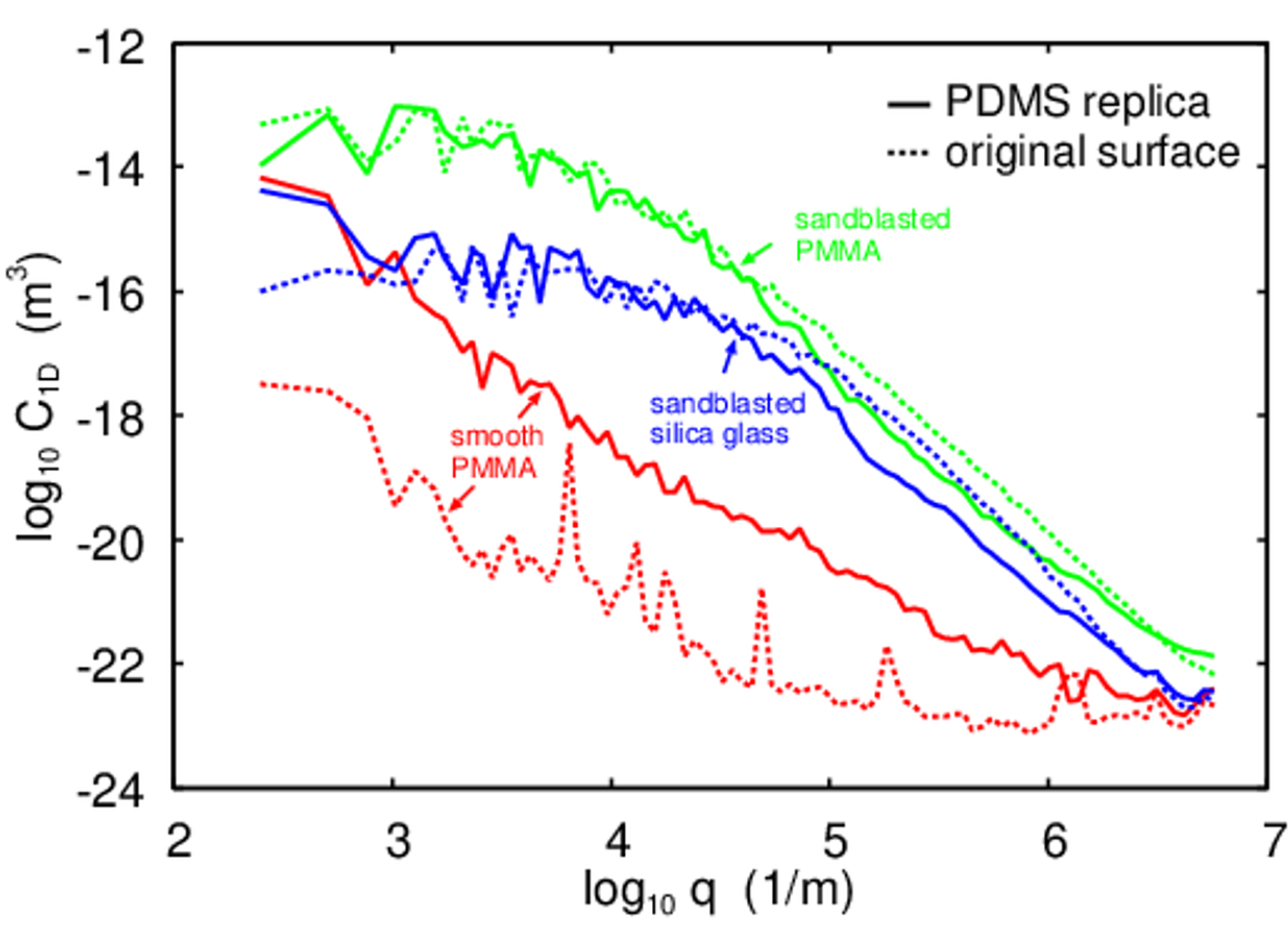}
\caption{\label{ADDED.PDMS.on.GLASS.and.PMMA.eps}
The surface roughness power spectra of PDMS molded against three different surfaces, smooth PMMA (red), sandblasted silica glass (blue), and sandblasted PMMA (green). The solid lines are the power spectra of the PDMS replica and the dashed lines are of the originals. The power spectra were obtained using the 
Surftest stylus instrument by averaging over two line scans each $25 \ {\rm mm}$ long.}
\end{figure}

\begin{figure}[!ht]
        \includegraphics[width=0.45\textwidth]{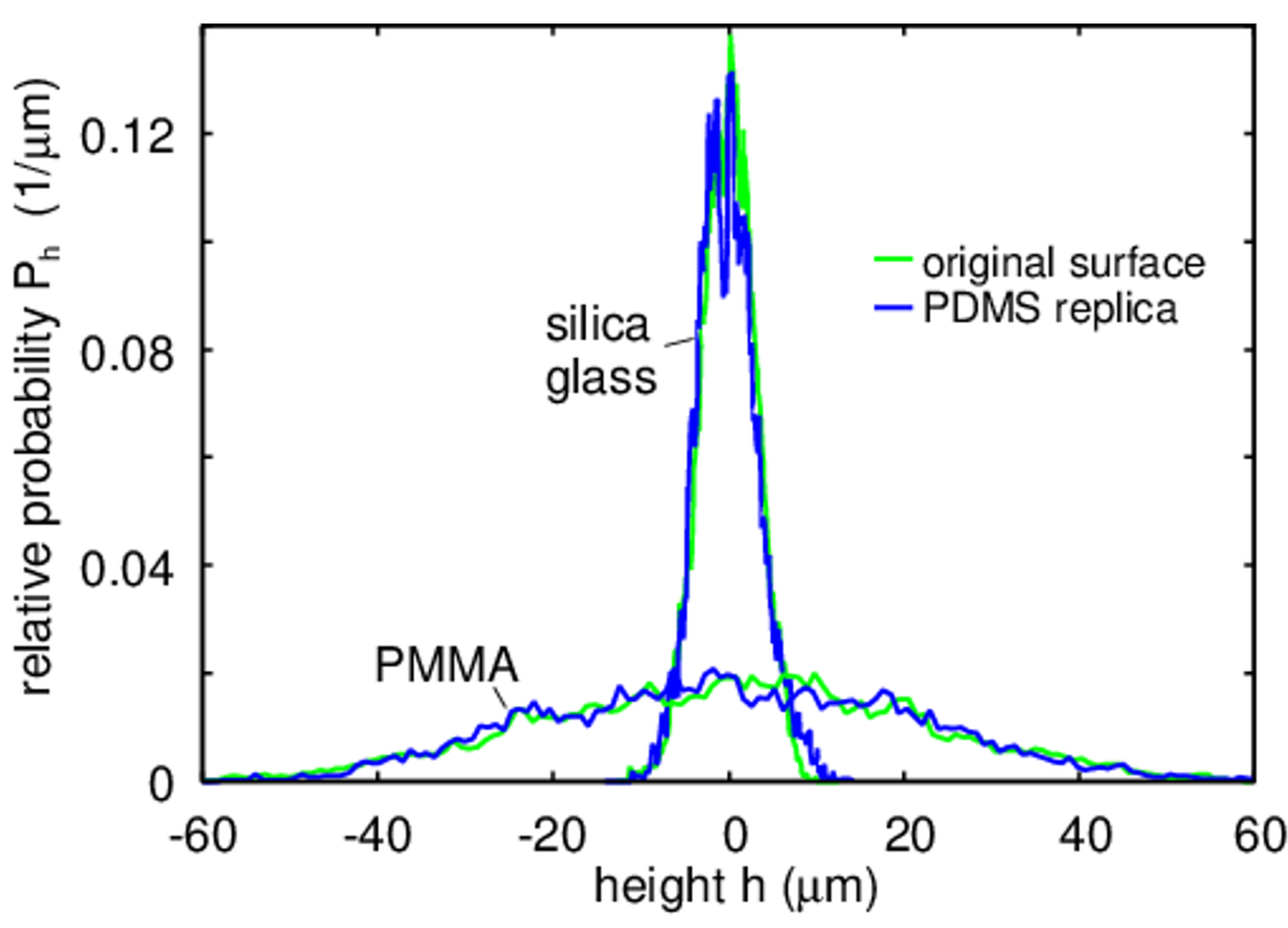}
\caption{\label{ADDED.1h.2.Ph.PDMS.replica.glass.PMMA.eps}
The height probability distribution of the sandblasted silica glass 
and PMMA surfaces (green lines) and corresponding PDMS replicas (blue).
}
\end{figure}

Fig. \ref{ADDED.PDMS.on.GLASS.and.PMMA.eps}
shows the surface roughness power spectra of PDMS molded against three different surfaces, 
namely smooth PMMA (red), sandblasted silica glass (blue), and sandblasted PMMA (green).
The power spectra were obtained using the Surftest stylus instrument by averaging over two line scans each $25 \ {\rm mm}$ long.
The solid lines are the power spectra of the PDMS replica and the dashed lines are of the originals. For the two sandblasted surfaces the PDMS replicas and the original surfaces give very similar power spectra, and in particular, the power spectra of the long wavelength roughness are well reproduced. This is also illustrated in Fig. \ref{ADDED.1h.2.Ph.PDMS.replica.glass.PMMA.eps} which shows the good agreement in height probability distribution of the sandblasted silica glass and PMMA surfaces (green lines) and the corresponding PDMS replicas (blue).

\begin{figure}[!ht]
        \includegraphics[width=0.45\textwidth]{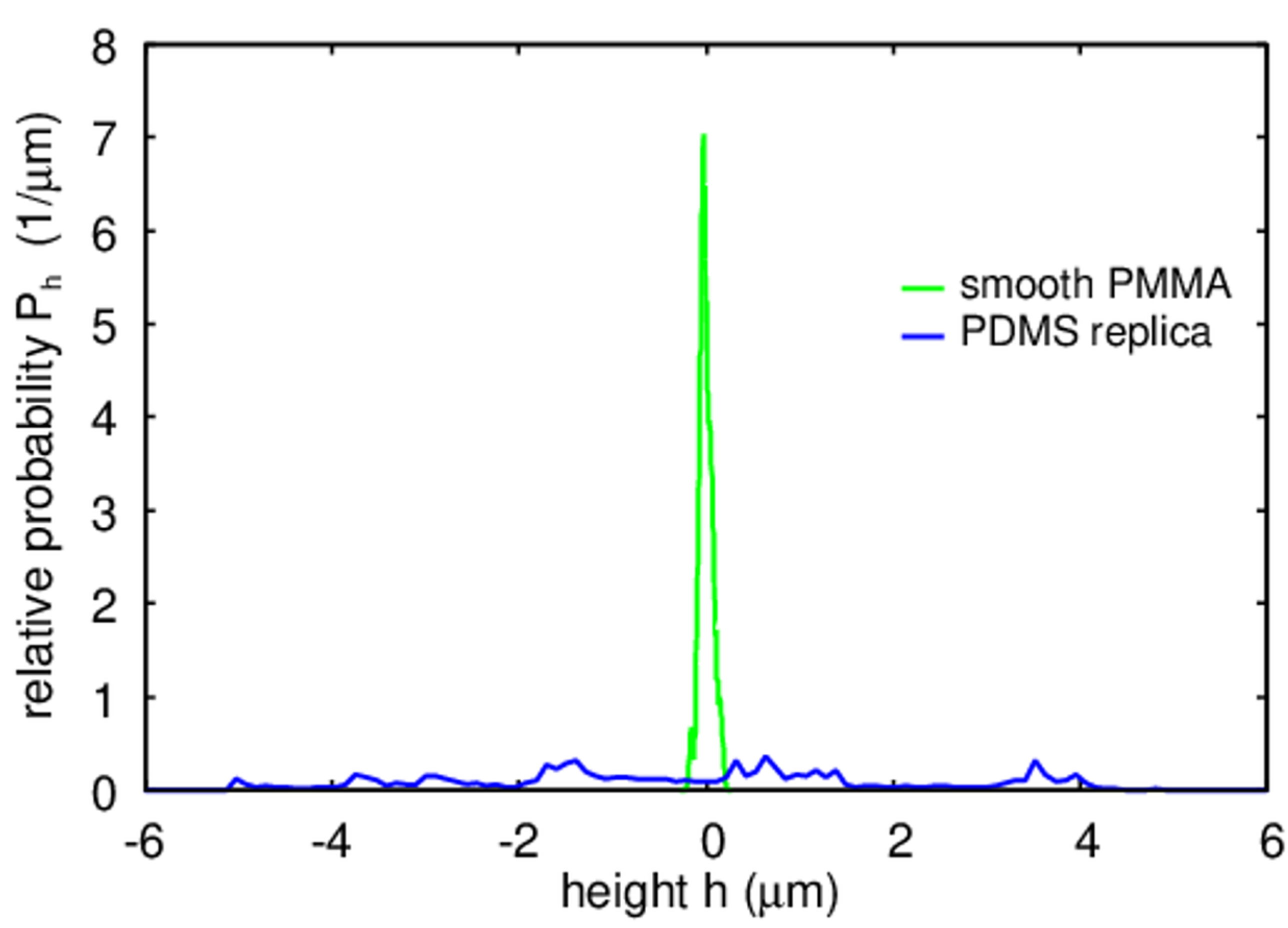}
\caption{\label{ADDED.1h.2.Ph.PDMS.replica.SMOOTH.PMMA.eps}
The height probability distribution of the smooth 
PMMA surface (green line) and the  PDMS replica of this surface (blue).
}
\end{figure}

It can be observed that the power spectra of the smooth PMMA surface and its PDMS replica differ by several orders of magnitude. Additionally, the obtained height distributions are entirely different (see Fig. \ref{ADDED.1h.2.Ph.PDMS.replica.SMOOTH.PMMA.eps}). This indicates that using engineering stylus measurements on soft rubbers with very smooth surfaces is likely to be unsuccessful. However, on surfaces with sufficient roughness, the stylus topography of the PDMS replica may be accurate. Nevertheless, considering our previous results for similar PDMS surfaces, there is no certainty that this will always be the case \cite{Julia}.

\vskip 0.3cm
{\bf 9 Discussion}

We have compared the power spectra obtained from surface topography measurements using different experimental instruments. We found that stylus instruments (AFM and two engineering stylus) give overlapping power spectra in the range of roughness length scales common to both techniques. The large modulus of the studied samples, the small tip diameter ($8 \ {\rm nm}$ for the AFM method), and the low contact forces applied in these two techniques result in accurate topography profiles and power spectra. For the surfaces studied, the power spectra obtained using optical instruments differed significantly from the results of the AFM and stylus studies. This is consistent with our earlier observations \cite{withL}.

We note, however, that for other types of materials and other types of roughness, the optical methods may give useful results. This is illustrated in Fig. \ref{1logq.2logC1D.lapping.all.eps} where we compare stylus and AFM data with the optical LIP method for a lapped nickel surface used as a roughness standard.

Studies based on optical methods are fast and give nice-looking topography pictures, but our study shows that in general the relative size of surface structures cannot be trusted. We recommend against using optical methods for quantitative surface roughness studies.

\vskip 0.3cm
{\bf 10 Summary and conclusion}

Analytical contact mechanics theories depend on the surface roughness via the surface roughness power spectrum. We have shown that height data obtained using optical methods are often inaccurate and should not be used for calculating surface roughness power spectra, while engineering stylus instruments and atomic force microscopy (AFM) generally give good results. For surfaces with isotropic roughness, all information about the roughness is contained in a line scan if it is long enough. For this case, we have shown how the 2D power spectrum $C_{\rm 2D}(q)$ can be obtained from the 1D power spectrum $C_{\rm 1D}(q)$
using different methods.

\vskip 0.3cm
{\bf Acknowledgments:}
We thank T.D.B. Jacobs, N. Miller, M.H. M\"user and L. Pastewka for the
samples used for the topography study.
We thank Cynthia Fuentes (BD Medical, Le Pont de Claix, France) and Mike
Pena (BD Medical, Franklin Lakes, NJ, US) for help with optical and
laser profilometry measurements, and Niko Cosa (Keyence, US) for help with white 
light profilometry and A.E. Yakini (MultiscaleConsulting and FZ J\"ulich)
for performing the Mitutoyo stylus measurements, and
Istiaque Choudhury and Nicholas  Strandwitz  (Materials Science and Engineering, at 
Lehigh University, Allentown, PA, US) for the XRR measurement and comments.

\vskip 0.3cm
{\bf Appendix A: Power spectra with roll-off region}

We assume that for both the 1D and 2D power spectra there is a flat roll-off region for
$q_0<q<q_{\rm r}$ and (self affine fractal) power-law behavior for $q_{\rm r} < q < q_1$.
We denote a roll-off region as flat when the power spectrum is constant.
We assume $q_0/q_{\rm r} << 1$ and $q_{\rm r}/q_1 <<1$ and $H>0$.
For this case if the 2D power spectrum
$$C_{\rm 2D} = C_0 \left ({q\over q_{\rm r}} \right )^{-2(1+H)}\eqno(A1)$$
for $q>q_{\rm r}$ and 
$$C_{\rm 2D} = C_0\eqno(A2)$$
for $q< q_{\rm r}$ then (10) gives
$$h_{\rm rms}^2 = 2\pi C_0 \int_{q_{\rm r}}^\infty dk k \ \left ({k\over q_{\rm r}} \right )^{-2(1+H)} 
+ 2 \pi C_0 \int_0^{q_{\rm r}} dk k$$
giving 
$$C_0 = {1 \over \pi}  h_{\rm rms}^2 {H\over 1+H}{1\over q_{\rm r}^2}$$

In a similar way to the 1D power spectrum
one can show that for $q_{\rm r} < q$ 
$$C_{\rm 1D} = {1+H \over 1+2H} \pi q_{\rm r} C_0 \left ( {q\over q_{\rm r}}\right )^{-1-2H}\eqno(A3)$$
and for $q<q_{\rm r}$
$$C_{\rm 1D} = {1+H \over 1+2H} \pi q_{\rm r} C_0 \eqno(A4)$$
so $C_{\rm 1D}$ is continuous for $q=q_{\rm r}$. 

Suppose we know that $C_{\rm 1D}$ is of the form,A4). In this case, if the 2D power spectrum is of the form (A1)-(A2), then in the self-affine fractal region $C_{\rm 2D} = \pi q C_{\rm 1D} (1+2H)/(1+H)$. This relation is not valid in the roll-off region and is not exact even for $q > q_{\rm r}$ as we will show below. This implies that if the 1D power spectrum has a flat roll-off region, this is not the case for the corresponding 2D power spectrum.

\begin{figure}[!ht]
        \includegraphics[width=0.45\textwidth]{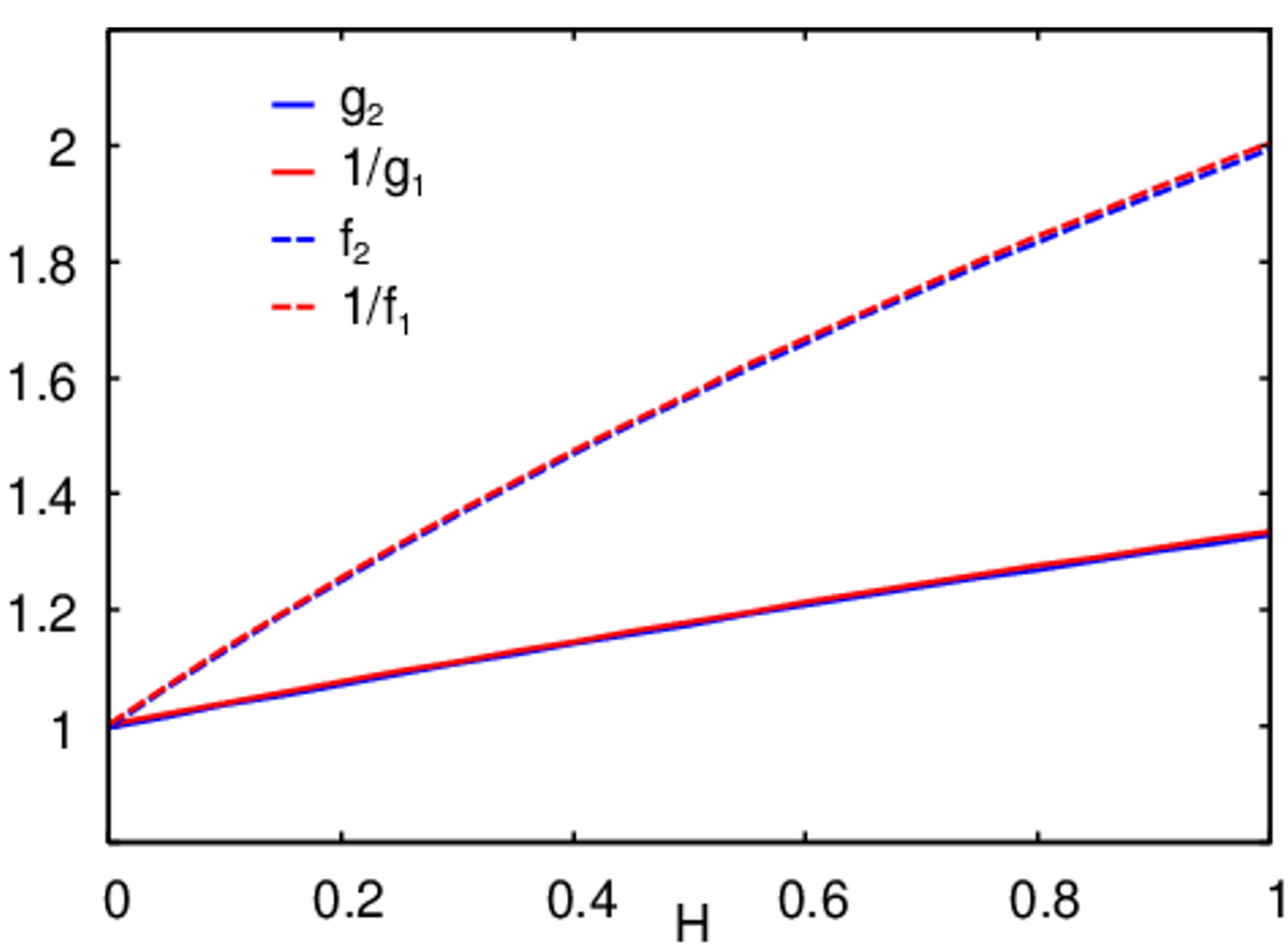}
\caption{\label{1H.2g.rollOFF.eps}
The 2D power spectrum factors $g_2(H)$ and $f_2(H)$, and the inverse $1/g_1(H)$ and $1/f_1(H)$ of the 1D power spectrum factors,
as a function of the Hurst exponent $H$.
}
\end{figure}

Assume that the 1D power spectrum is given by (A3)-(A4).
Substituting (A3) in (13) and writing $k=x q_{\rm r}$ 
gives for $q>q_{\rm r}$ the 2D power spectrum
$$C_{\rm 2D} = C_0 (1+H) \left ({q\over q_{\rm r}}\right )^{-2(1+H)}
\int_1^\infty dx \ {x^{-2-2H} \over (x^2-1)^{1/2}}\eqno(A5)$$
and for $q<q_{\rm r}$ we get
$$C_{\rm 2D} = C_0 (1+H) \int_1^\infty dx \ {x^{-2-2H} \over (x^2-(q/q_{\rm r})^2)^{1/2}}\eqno(A6)$$
For $q/q_{\rm r} <<1 $ this gives
$$C_{\rm 2D} = C_0 (1+H) \int_1^\infty dx \ x^{-3-2H} =  {1\over 2} C_0$$
For $q>q_{\rm r}$ we write
$$C_{\rm 2D} = C_0 \left ({q\over q_{\rm r}}\right )^{-2-2H} g_2(H)\eqno(A7)$$
where
$$g_2(H) = (1+H)\int_1^\infty dx \ {x^{-2-2H} \over (x^2-1)^{1/2}}\eqno(A8)$$
In Fig. \ref{1H.2g.rollOFF.eps} we show $g_2(H)$.

        \begin{figure}[!ht]
        \includegraphics[width=0.45\textwidth]{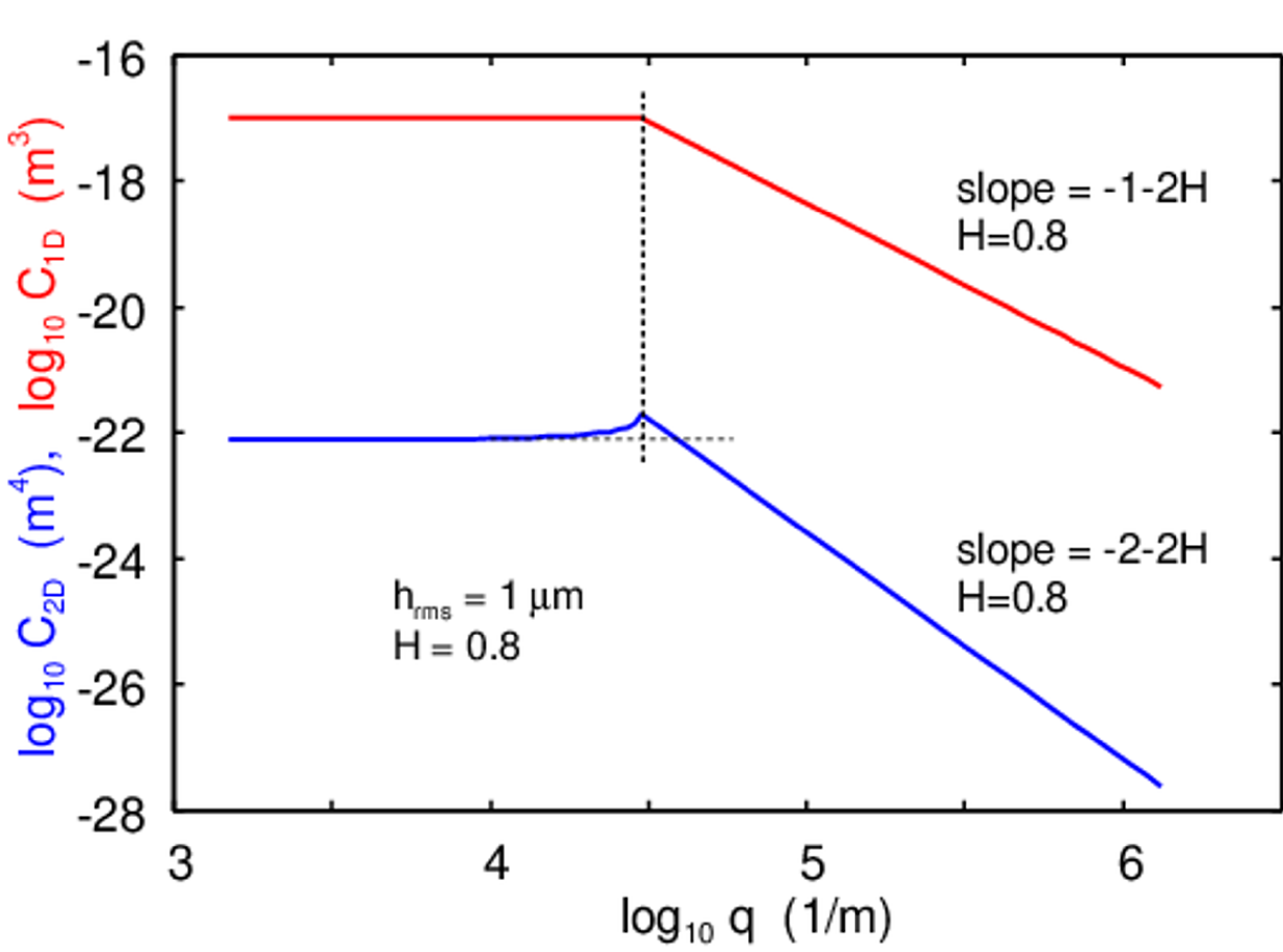}
\caption{\label{1logq.2logC.math.givenC1D.eps}
The 1D (red line) and 2D (blue line) power spectra for a randomly rough surface with
the rms roughness $h_{\rm rms} = 1 \ {\rm \mu m}$ and Hurst exponent $H=0.8$. 
The 1D power spectrum has a flat roll-off region and the 2D power spectrum is calculated from
$C_{\rm 1D}$ using (13). The small, large and
roll-off wavenumbers are $q_0=1\times 10^3 \ {\rm m}^{-1}$, $q_1=2048 \times 10^3 \ {\rm m}^{-1}$
and $q_{\rm r}=30 \times 10^3 \ {\rm m}^{-1}$, respectively.
}
\end{figure}

        \begin{figure}[!ht]
        \includegraphics[width=0.45\textwidth]{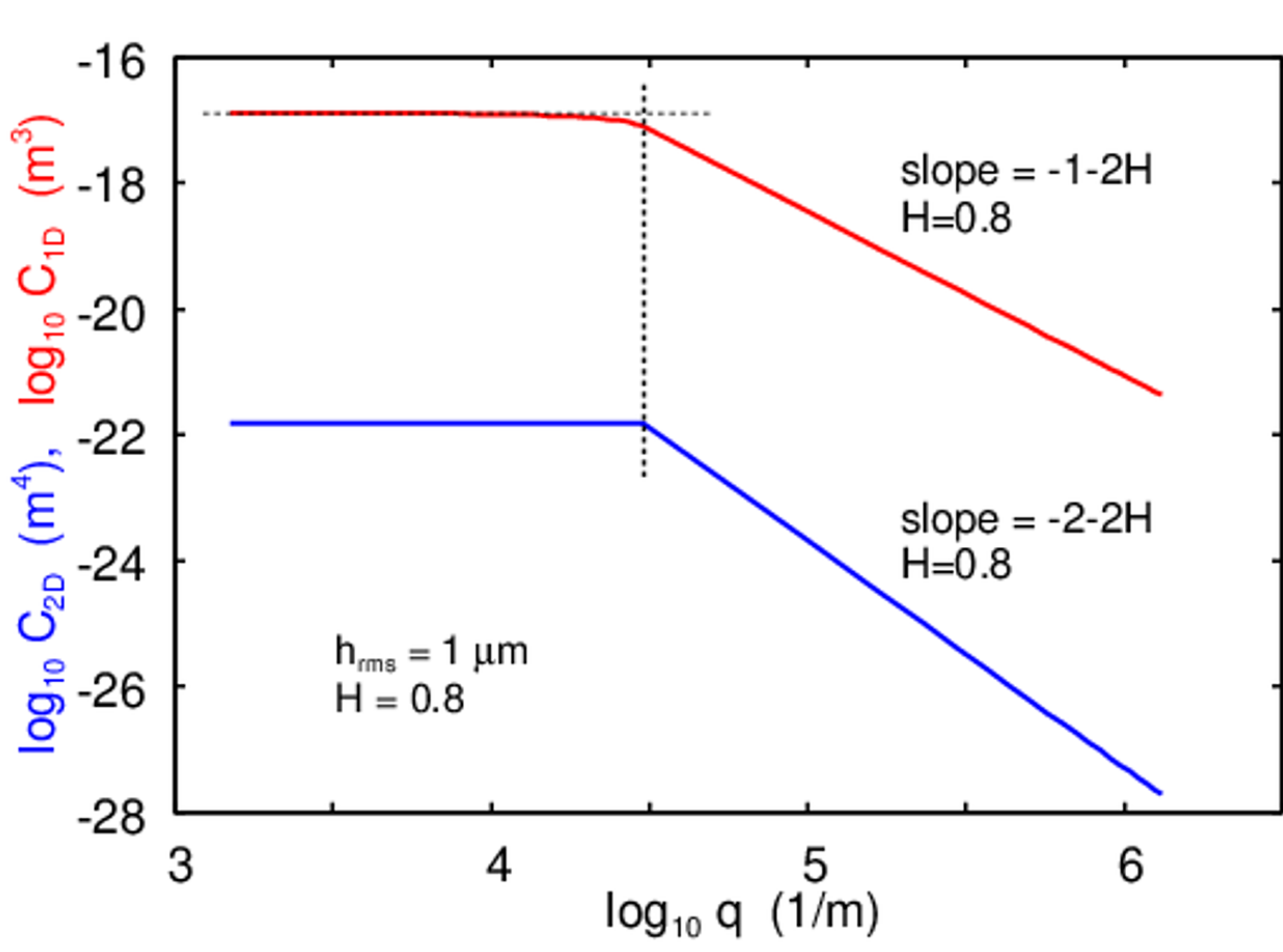}
\caption{\label{1logq.2logC.C2given.eps}
The 1D (red line) and 2D (blue line) power spectra for a randomly rough surface with
the rms roughness $h_{\rm rms} = 1 \ {\rm \mu m}$ and Hurst exponent $H=0.8$. 
The 2D power spectrum has a flat roll-off region and the 1D power spectrum is calculated from
$C_{\rm 1D}$ using (14). The small, large and
roll-off wavenumbers are $q_0=1\times 10^3 \ {\rm m}^{-1}$, $q_1=2048 \times 10^3 \ {\rm m}^{-1}$
and $q_{\rm r}=30 \times 10^3 \ {\rm m}^{-1}$, respectively.
}
\end{figure}

The analysis above shows that if $C_{\rm 1D}$ has a self-affine region and a flat roll-off region then
$C_{\rm 2D}$ will have a self-affine region which differs by a constant factor
$g_2(H)$ from what one would expect if the roll-off region of $C_{\rm 2D}$ would be flat 
[compare (A1) with (A7)]. Indeed the roll-off region close to $q_{\rm r}$ [as given by (A6)] 
is not perfectly flat as shown in Fig. \ref{1logq.2logC.math.givenC1D.eps}. 
We note that the exact form of the roll-off region is usually not very
important and for most purposes the 2D power spectrum can be obtained from (A7) with $q_{\rm r}$, $h_{\rm rms}$ and $H$ 
determined from the 1D power spectrum. This is equivalent to using the relation
$C_{\rm 2D}=f_2(H)C_{\rm 1D}/(\pi q)$ with $f_2(H)=g_2(H)(1+2H)/(1+H)$ and a constant in the roll-off region given by
$C_{\rm 2D}(q_{\rm r})$.

In Sec. 3 we studied the power spectrum of a rough surface which was generated using the power spectrum
with a 2D power spectrum with a flat roll-off. 
In this case, the 1D power spectrum does not have a perfectly flat roll-off region but instead shows a region close to $q_{\rm r}$ with reduced magnitude
(see Fig. \ref{1logq.2logC.red1D.blue2D.eps}). 
This result also follows the related relation (14) as we now will show.

Assume that the 2D power spectrum is given by (A1)-(A2). 
Using (A1) and (A2) in (14) and writing $k=x q_{\rm r}$ 
gives for $q>q_{\rm r}$ the 1D power spectrum
$$C_{\rm 1D} = 2C^0 q_{\rm r} \left ({q\over q_{\rm r}}\right )^{-1-2H} 
\int_1^\infty dx \ {x^{-1-2H} \over (x^2-1)^{1/2}}\eqno(A9)$$
and for $q<q_{\rm r}$ we get
$$C_{\rm 1D} = 2C^0 \left [ \left (q_{\rm r}^2 -q^2 \right )^{1/2} 
+ q_{\rm r} \int_1^\infty dx \ {x^{-1-2H} \over (x^2-(q/q_{\rm r})^2)^{1/2}}\right ] \eqno(A10)$$
For $q/q_{\rm r} <<1 $ this gives
$$C_{\rm 1D} = {4\over \pi} {1+H \over 1+2H} \pi q_{\rm r} C_0 \eqno(A11)$$
The analysis above shows that if $C_{\rm 2D}$ has a self-affine region and a flat roll-off region then
$C_{\rm 1D}$ will have a self-affine region that differs by a constant factor
$g_1(H)$ from what one would expect if the roll-off region of $C_{\rm 1D}$ would be flat 
[compare (A3) with (A11)]. Indeed the roll-off region close to $q_{\rm r}$ [as given by (A11)] 
is not perfectly flat as shown in Fig. \ref{1logq.2logC.C2given.eps}. 

For $q>q_{\rm r}$ we write
$$C_{\rm 1D} = {1+H \over 1+2H} \pi q_{\rm r} C_0 \left ( {q\over q_{\rm r}}\right )^{-1-2H}g_1(H)\eqno(A12)$$
where
$$g_1 ={2\over \pi} {1+2H\over 1+H} \int_1^\infty dx \ {x^{-1-2H} \over (x^2-1)^{1/2}}\eqno(A13)$$
We write $C_{\rm 1D}=f_1(H) \pi q C_{\rm 2D}$ with $f_1(H)=g_1(H)(1+H)/(1+2H)$ 
and a constant in the roll-off region given by $C_{\rm 1D}(q_{\rm r})$.
In Fig. \ref{1H.2g.rollOFF.eps} we show $1/g_1 (H)$ and $1/f_1(H)$. Note that $1/g_1 \approx g_2$ and 
$1/f_1 \approx f_2 \approx (1+3H)^{1/2}$
Hence we can use the relation $C_{\rm 2D}=f(H) C_{\rm 1D}/(\pi q)$
with $f(H) \approx (1+3H)^{1/2}$ to 
calculate $C_{\rm 2D}$ in the self-affine fractal region when $C_{\rm 1D}$ is given
(with a flat roll-off region), and also to 
calculate $C_{\rm 1D}$ if the self-affine fractal region when $C_{\rm 2D}$ is given
(with a flat roll-off region). In practical applications only the first application is relevant.


\begin{thebibliography}{99}
\bibitem{WeeDefined}
On a Bruker Dektak XT info page, one can read:

``The power of proper
filtering for data analysis, according to recognized ISO
standard methods, cannot be overestimated when striving
to provide the most accurate and reproducible results for a
measuring system'' and 
``The short cutoff filter with a cutoff
spatial frequency of $\lambda_{\rm s}$ is applied to this total profile
in order to produce the ``primary profile''. The spatial
frequencies that are rejected by the short cutoff are
considered noise ... '' 

Filters are used for several purposes: (a) to make some
quantities, like the rms slope, more well-defined or (b) to
eliminate undesired vibrations of the tip during the scan. 
(See, ISO-Standardized Filtering for DektakXT Stylus Profilers, Application Note
550, Matt Novak and Son Bui, Bruker Nano Surfaces Division, Tucson, AZ USA.)
However, it is often possible to avoid tip vibration by tuning the
contact force and scan speed. In general, filtering should never be used if the topography
is used as input for theory calculations.
\end{thebibliography}
\end{document}